\documentclass[aps,prd,twocolumn,superscriptaddress,showpacs,showkeys]{revtex4}

\usepackage{epsfig,epsf}
\usepackage{amsmath}
\usepackage{amsthm}
\usepackage{amsfonts}
\usepackage{amssymb}
\usepackage{dsfont}

\usepackage{multirow}

\usepackage{slashed}

\usepackage[active]{srcltx}

\def\OMIT#1{}

\newcommand{\be}{\begin{equation}}
\newcommand{\ee}{\end{equation}}

\DeclareMathOperator{\Tr}{Tr}

\newcommand \vev [1] {\langle{#1}\rangle}

\newcommand \ket [1] {|{#1}\rangle}
\newcommand \bra [1] {\langle {#1}|}

\begin{document}

%%%%%%%%%%%%%%%%%%%%%%%%%%%%%%%%%%%%%%%%%%
%Define Title, Author, Address, Preprint#

\title{Higher twist parton distributions from light-cone wave functions}

\date{\today}

\author{V.M.~Braun}
\affiliation{Institut f\"ur Theoretische Physik, Universit\"at
   Regensburg,D-93040 Regensburg, Germany}
\author{T.~Lautenschlager}
\affiliation{Institut f\"ur Theoretische Physik, Universit\"at
   Regensburg,D-93040 Regensburg, Germany}
\author{A.N.~Manashov}
\affiliation{Institut f\"ur Theoretische Physik, Universit\"at
   Regensburg,D-93040 Regensburg, Germany}
\affiliation{Department of Theoretical Physics,  St.-Petersburg State
University 199034, St.-Petersburg, Russia}
\author{B.~Pirnay}
\affiliation{Institut f\"ur Theoretische Physik, Universit\"at
   Regensburg,D-93040 Regensburg, Germany}
\date{\today}

\begin{abstract}
  \vspace*{0.3cm}
We explore the possibility to construct higher-twist parton distributions
in a nucleon at some low reference scale
from convolution integrals of the light-cone wave functions (WFs).
To this end we introduce simple models for the four-particle
nucleon WFs involving three valence quarks and a gluon with
total orbital momentum zero, and estimate their normalization
(WF at the origin) using QCD sum rules. We demonstrate that
these WFs provide one with a reasonable description of both
polarized and unpolarized gluon parton densities at large values of
Bjorken variable $x\ge 0.5$. Twist-three parton distributions are
then constructed as convolution integrals of $qqqg$ and usual
three-quark WFs. The cases of the polarized structure
function $g_2(x,Q^2)$ and single transverse spin asymmetries are considered
in detail. We find that the so-called gluon-pole contribution
to twist-three distributions relevant for single
spin asymmetry vanishes in this model, but is generated
perturbatively at higher scales by the evolution, in the spirit of GRV parton
distributions.
 \end{abstract}



\maketitle

%
%%%%%%%%%%%%%%%%%%%%%%%%%%%%%%%%%%%%%%%%%%
\section{Introduction}
%%%%%%%%%%%%%%%%%%%%%%%%%%%%%%%%%%%%%%%%%%%
%

Higher-twist parton distributions are conceptually very interesting as they go beyond
the simple parton model description and allow one to quantify correlations between
the partons. Unfortunately, they prove to be very elusive. Despite considerable efforts,
very little is known even about the simplest, twist-three distributions which contribute,
e.g. structure function $g_2(x,Q^2)$ in the polarized deep-inelastic scattering
\cite{Efremov:1983eb,Bukhvostov:1984as,Ratcliffe:1985mp,Balitsky:1987bk,Balitsky:1989jb,Ali:1991em,Kodaira:1996md,Kodaira:1998jn,Accardi:2009au}
and transverse single spin asymmetries (SSAs) in the collinear factorization approach
\cite{Efremov:1981sh,Efremov:1984ip,Qiu:1991pp,Qiu:1991wg,Efremov:1994dg,Qiu:1998ia,Kanazawa:2000hz,Kanazawa:2000cx,Eguchi:2006mc,Koike:2007rq,Kang:2008ih}.

One general reason for this is that the structure of higher twist parton distributions is
much more complicated compared to the leading twist: they are functions of two and more
parton momentum fractions.  The usual strategy to extract parton distributions
from experimental data has been to assume a certain functional form with a few
adjustable parameters at a reference scale, and find the parameters  by making global
fits to the available data. This is a standard approach which works quite well
for the leading twist. Unfortunately, it does not work for higher twist
(or, at least, has not been applied systematically) because
there is no physical intuition on how such distributions may look like.
Also the asymptotic behavior of higher-twist distributions
both at small and large $x$ is poorly understood.
Hence it is very hard to guess an adequate parametrization.

In this work we make a step in this direction. Recall that the case of
higher-twist parton distributions is not unique in that they are functions of several
kinematic variables: in studies of generalized parton distributions (GPDs)
or ``unintegrated'' transverse-momentum dependent distributions (TMDs)
the same complication arises. In both cases, representations in terms of overlap
integrals of light-cone wave functions have been extremely useful for developing the
underlying physics picture and provide one with a good basis for theoretical modelling.
In what follows we try to follow the same path for the construction of higher-twist
distributions as overlap integrals between Fock states with the minimum (valence) and
next-to-minimum (one extra gluon) parton content.

In order to keep the model as simple as possible, in this work we restrict ourselves
to contributions of the states with total zero angular momentum. We overtake
the expressions for three-quark wave functions from Ref.~\cite{Diehl:1998kh} which
have been shown \cite{Bolz:1996sw,Diehl:1998kh} to provide one with a good description  for
quark parton densities at large $x$ and the nucleon magnetic form factor.
The new contribution of this paper is to include into consideration the Fock states
with one additional gluon which were considered in~\cite{Diehl:1998kh}
on a qualitative level. We find that there exist three independent $qqqg$
wave functions with zero orbital momentum. Our analysis of their symmetry properties
does not agree with earlier results~\cite{Ji:2003yj}.
We calculate the normalization of these new wave functions using QCD sum rule approach and
construct explicit models by the requirement that their light-cone limit (zero transverse
separation) reproduces the nucleon twist-4 distribution amplitudes introduced in
Ref.~\cite{Braun:2008ia}.

Having specified the wave functions, we calculate the quark and gluon polarized and
unpolarized parton distributions and find agreement with the existing parametrizations
at large $x$ without any fine-tuning of the parameters.
Encouraged by this, we construct the twist-three correlation function
involving a quark, antiquark and gluon fields which is relevant for the
structure function $g_2(x,Q^2)$ and single spin asymmetries. In our model
this correlation function vanishes at the boundaries of parton regions
where one of the momentum fractions goes to zero, but non-zero values
are obtained at higher scales perturbatively through the QCD evolution.
This phenomenon is in full analogy to the generation of a large gluon parton distribution
at small $x$ starting from the ``valence''-like ansatz in the GRV approach~\cite{Gluck:1994uf,Gluck:1998xa}.
Such, radiatively generated, soft-gluon pole and soft-fermion pole contributions to the spin asymmetries
are calculated and compared to the existing parametrizations. The sign of radiatively generated
soft pole terms as well as the sign of the twist-three contribution to the structure
function $g_2(x,Q^2)$ at large $x$ are largely model-independent predictions of our approach;
these signs turn out to be in agreement with the data in all cases.
Finally, we discuss possible generalizations of our simple model that may
provide one with usable parametrizations for the phenomenological analysis.

%
%%%%%%%%%%%%%%%%%%%%%%%%%%%%%%%%%%%%%%%%%%%%%%%%%%%%%%%%%%%%%%%%%%%%%%%%%%%%%%%%%%%%%%%%%%%%%%%%%
\section{Light-cone coordinates}
%%%%%%%%%%%%%%%%%%%%%%%%%%%%%%%%%%%%%%%%%%%%%%%%%%%%%%%%%%%%%%%%%%%%%%%%%%%%%%%%%%%%%%%%%%%%%%%%%
%

For an arbitrary four-vector $a^\mu$ we define the light-cone coordinates as
\begin{eqnarray}
&& a_+ = \frac{1}{\sqrt{2}}(a^0+a^3)\,,\qquad
 a_- = \frac{1}{\sqrt{2}}(a^0-a^3)\,,
\nonumber\\
&& a = a^1 + i a^2\,, \hspace*{1.65cm}
\bar a = a^1-i a^2\,,
\end{eqnarray}
so that the matrix $a=a_\mu \sigma^\mu$, where $\sigma^\mu=(\mathbb{I},\vec{\sigma})$ takes the
form
\begin{align}\label{a}
a_{\alpha\dot\alpha}=a_\mu \sigma^\mu_{\alpha\dot\alpha}=\begin{pmatrix}\sqrt{2}a_-& -\bar a\\
-a &\sqrt{2}a_+\end{pmatrix}.
\end{align}
In what follows we use Weyl representation for
the $\gamma-$matrices
\begin{eqnarray}
&&\gamma^0= \begin{pmatrix} 0&\mathbb{I}\\\mathbb{I}&0\end{pmatrix}\,,\qquad
\gamma^i= \begin{pmatrix} 0&\sigma^i\\-\sigma^i&0\end{pmatrix}\,,
\nonumber\\
&&\gamma^5\equiv \gamma^0\gamma^1\gamma^2\gamma^3 =
\begin{pmatrix} -\mathbb{I}& 0 \\ 0& \mathbb{I}\end{pmatrix}\,
\end{eqnarray}
and the two-component notation for Dirac spinors
\begin{align}
   q = \begin{pmatrix}\psi_\alpha \\ \bar\chi^{\dot\alpha} \end{pmatrix}\equiv
 \begin{pmatrix}q_\downarrow\\ q_\uparrow \end{pmatrix}
   \,,\quad
\bar q = q^\dagger\gamma^0 = (\chi^\alpha,\bar\psi_{\dot \alpha}) \equiv (\bar q_\downarrow,\bar q_\uparrow)\,.
\end{align}
The two independent light-like vectors
\begin{align}
  n^\mu = \frac{1}{\sqrt{2}}(1,0,0,-1)\,, \qquad \tilde n^\mu = \frac{1}{\sqrt{2}}(1,0,0,1)\,,
\end{align}
$n^2=\tilde n^2=0$, $n\tilde n=1$ can be parametrized
in terms of the two auxiliary Weyl spinors:
\begin{equation}
n_{\alpha\dot\alpha}=\lambda_\alpha\bar \lambda_{\dot\alpha}\,,\qquad
\tilde n_{\alpha\dot\alpha}=\mu_\alpha\bar \mu_{\dot\alpha}\,,
\end{equation}
where
\begin{eqnarray}\label{lambdamu}
\lambda_\alpha=2^{1/4}\begin{pmatrix}-1 \\0\end{pmatrix},&&
\mu_\alpha=2^{1/4}\begin{pmatrix}0 \\1\end{pmatrix},
\nonumber\\
\bar\lambda_{\dot\alpha}=2^{1/4}\begin{pmatrix}-1 \\ 0 \end{pmatrix},&&
\bar\mu_{\dot\alpha}=2^{1/4}\begin{pmatrix}0  \\ 1\end{pmatrix}.
\end{eqnarray}
We accept the following rules for raising and lowering the spinor
indices~(cf. Ref.~\cite{Braun:2008ia})
\begin{equation}
\lambda^\alpha=\epsilon^{\alpha\beta}\lambda_\beta\,,\quad
\lambda_\alpha=\lambda^\beta\epsilon_{\beta\alpha}\,,\quad
\bar\lambda^{\dot\alpha}=\bar \lambda_{\dot\beta}\epsilon^{\dot\beta\dot\alpha}\!\!, \quad
\bar \lambda_{\dot\alpha}=\epsilon_{\dot\alpha\dot\beta}\bar \lambda^{\dot\beta}\!,
\end{equation}
where the antisymmetric Levi-Civita tensor is defined as
$$
\epsilon_{12}=\epsilon^{12}=
-\epsilon_{\dot 1\dot 2}=-\epsilon^{\dot 1\dot 2}=1\,.
$$
The auxiliary spinors $\lambda$ and $\mu$ are normalized as
\begin{align}
 (\mu\lambda) = \mu^\alpha\lambda_\alpha = -(\lambda\mu) = - \sqrt{2}\,,
\nonumber\\
 (\bar\mu\bar\lambda) =
  \bar\mu_{\dot\alpha}\bar\lambda^{\dot\alpha} = -(\bar\lambda\bar \mu) = + \sqrt{2}\,
\end{align}
and serve to specify "plus" and "minus" components of the fields.
We define
\begin{eqnarray}
\psi_+=\lambda^\alpha \psi_\alpha, &&
\psi_-=\mu^\alpha \psi_\alpha,
\nonumber\\
\bar \chi_+= \bar\chi_{\dot\alpha}\bar\lambda^{\dot\alpha},&&
\bar \chi_-=  \bar\chi_{\dot\alpha}\bar\mu^{\dot\alpha} \,,
\end{eqnarray}
so that each two-component spinor can be decomposed as
\begin{align}
 (\mu\lambda) \psi_\alpha = \lambda_\alpha \,\psi_- - \mu_\alpha\, \psi_+\,,
\nonumber\\
 (\bar\lambda\bar\mu) \bar\chi^{\dot\alpha} =
 \bar\lambda^{\dot\alpha}\, \bar\chi_- - \bar\mu^{\dot\alpha}\, \bar\chi_+\,.
\end{align}
In the same notation the light-cone decomposition of a vector (e.g. gluon) field
takes the form
\begin{align}
  A_{\alpha\dot\alpha} &=  A_- \,\lambda_\alpha\bar\lambda_{\dot\alpha}
                   + A_+ \,\mu_\alpha\bar\mu_{\dot\alpha}
                   + \frac{\bar A}{\sqrt{2}} \, \lambda_\alpha \bar\mu_{\dot\alpha}
                   + \frac{A}{\sqrt{2}}\, \mu_\alpha \bar\lambda_{\dot\alpha}\,.
\label{vectordecompose}
\end{align}
The "plus" spinor fields $\psi_+,\bar\chi_+$ and transverse gluon fields $A,\bar A$
are assumed to be the dynamical fields in the light-cone quantization framework.
The  "minus" fields $\psi_-,\bar\chi_-, A_-$ can be expressed in terms of the dynamical ones
with the help of equations of motion (EOM) whereas $A_+=0$ due to the gauge fixing condition.

The plus quark fields have the following canonical expansion
\begin{align}\label{qupdown}
q_{\downarrow+}(x)=&
\int\! \frac{dp_+}{\sqrt{2p_+}}  \frac{d^2p_\perp}{(2\pi)^3} \theta(p_+)
\biggl[ e^{-ipx} b_\downarrow(p)+e^{+ipx} d_\uparrow^\dagger(p)\biggr],
\notag\\
q_{\uparrow+}(x)=&\int\! \frac{dp_+}{\sqrt{2p_+}}  \frac{d^2p_\perp}{(2\pi)^3} \theta(p_+)
\biggl[ e^{-ipx} b_\uparrow(p)+e^{+ipx} d^\dagger_\downarrow(p)\biggr],
\end{align}
where $b_{\uparrow (\downarrow)},d_{\uparrow (\downarrow)}$ are the annihilation operators
of quark and antiquark of positive (negative) helicity, respectively. They obey the
standard anticommutation relations
\begin{eqnarray}
\lefteqn{\hspace*{-0.5cm}
\{b_{\lambda}(p),b_{\lambda'}^\dagger(p')\}=\{d_{\lambda}(p),d^\dagger_{\lambda'}(p')\}=}
\nonumber\\&=&
2p_+ (2\pi)^3\delta_{\lambda,\lambda'}\delta(p_+-p'_+)\delta^{2}(p_\perp-p'_\perp)\,.
\end{eqnarray}
The similar expansion for the dynamical transversely polarized
gluon fields $A$ and $\bar A$ reads
\begin{eqnarray}
\bar A(x)&\!=\!&\sqrt{2}\!\int\! \frac{dk_+}{2k_+}  \frac{d^2k_\perp}{(2\pi)^3} \theta(k_+)
\biggl[ e^{-ikx} a_\uparrow (k)+e^{+ikx} a_\downarrow^\dagger(k)
\biggr],
\nonumber\\
A(x)&\!=\!&\sqrt{2}\!\int\! \frac{dk_+}{2k_+}  \frac{d^2k_\perp}{(2\pi)^3} \theta(k_+)
\biggl[ e^{-ikx} a_\downarrow (k)+e^{+ikx} a_\uparrow^\dagger(k)\biggr].
\nonumber\\
\label{a+a}
\end{eqnarray}
Here and below $A=\sum_a t^a A^a$ etc. where $t^a$ are the usual $SU(3)$ generators in the
fundamental representation, normalized as $tr(t^a t^b)=\dfrac12 \delta^{ab}$.
The creation and annihilation operators obey the commutation relation
\begin{eqnarray}
\lefteqn{\Big[a^b_\lambda(p),(a^{b'}_{\lambda'}(p'))^\dagger\Big]=}
\nonumber\\&=&
2p_+ (2\pi)^3\delta_{\lambda,\lambda'}\delta^{bb'}\delta(p_+-p'_+)\delta^{2}(p_\perp-p'_\perp)\,.
\end{eqnarray}
{}Finally, the gluon strength tensor $F_{\mu\nu}$ and its dual
$\widetilde{F}_{\mu\nu}$ can be decomposed as
\begin{align}
F_{\alpha\beta,\dot\alpha\dot\beta}&=\sigma^\mu_{\alpha\dot\alpha}\sigma^\nu_{\beta\dot\beta}
F_{\mu\nu}=
2\Big(\epsilon_{\dot\alpha\dot\beta} f_{\alpha\beta}-
\epsilon_{\alpha\beta} \bar f_{\dot\alpha\dot\beta}
\Big)\,,
\notag\\
i {\widetilde F}_{\alpha\beta,\dot\alpha\dot\beta}&=2\Big(\epsilon_{\dot\alpha\dot\beta}
f_{\alpha\beta}+
\epsilon_{\alpha\beta}\bar f_{\dot\alpha\dot\beta}\Big)\,.
\end{align}
Here $f_{\alpha\beta}$ and $\bar f_{\dot\alpha\dot\beta}$ are chiral and antichiral
symmetric tensors, $f_{\alpha\beta}= f_{\beta\alpha}$,
$\bar f = f^*$, which belong to $(1,0)$ and $(0,1)$ representations
of the Lorenz group, respectively.
Their ''good components'' are defined as
\begin{align}
f_{++}=\lambda^\alpha\lambda^\beta f_{\alpha\beta}\,,
&&
\bar f_{++}=\bar\lambda^{\dot\alpha}\bar\lambda^{\dot\beta} \bar f_{\dot\alpha\dot\beta}\,.
\end{align}
In the light-cone gauge
\begin{equation}
f_{++}=- \partial_+ A\,,\qquad \bar f_{++}=-\partial_+ \bar A\,,
\end{equation}
where $\partial_+ = n^\mu\partial_\mu = \partial/\partial x_-$,
so that they can readily be expanded in contributions of  annihilation and creation
operators using Eq.~(\ref{a+a}).

As mentioned above, "minus" field components can be expressed in terms of the
dynamical fields using QCD equations of motion.

%%%%%%%%%%%%%%%%%%%%%%%%%%%%%%%%%%%%%%%%%%%%%%%%%%%%%%%%%%%%%%%%%%%%%%%%%%%%%%%%%%%%%%%%%%%%%%
\section{Nucleon light-cone wave functions}\label{sect:light-cone}
%%%%%%%%%%%%%%%%%%%%%%%%%%%%%%%%%%%%%%%%%%%%%%%%%%%%%%%%%%%%%%%%%%%%%%%%%%%%%%%%%%%%%%%%%%%%%%

\subsection{Definitions and symmetry properties}

The light-cone wave functions (LCWFs) are defined as probability amplitudes
of the corresponding parton states which build up the proton with a given helicity.
They depend on parton longitudinal momentum fractions $x_i$, transverse momenta $k_{\perp i}$,
and on parton helicities. LCWFs are usually thought of as solutions of the eigenvalue problem
for the light-cone quantized QCD Hamiltonian~\cite{Kogut:1969xa,Brodsky:1997de},
although this construction is far from being complete.

Throughout this work we adopt some definitions and
partially also the notation from Ref.~\cite{Diehl:1998kh}.
In particular we use a shorthand notation for the $N$-parton differential phase space
\begin{align}\label{measurek}
[dx]_N=&\prod_{i=1}^N dx_i\,\delta(1-\sum x_i)\,,\notag
\\
[dk_\perp]_N=&\frac1{(16\pi^3)^{N-1}}\prod_{i=1}^N d^2k_{\perp,i}\,\delta^2\left(\sum k_{\perp i}\right)
\end{align}
and
\begin{align}
[\mathcal{D}X]_N=\frac{1}{\sqrt{x_1\ldots x_N}}[dx]_N[dk_\perp]_N\,.
\end{align}

The valence three-quark state  with zero angular momentum is the simplest one.
It can be described in terms of the single LCWF~\cite{Lepage:1980fj,Diehl:1998kh}
%\begin{widetext}
\begin{eqnarray}\label{Ansatz}
\ket{p,+}_{uud}%^{L_z=0}
&=&-\frac{\epsilon^{ijk}}{\sqrt{6}}\int[\mathcal{D}X]_3 \Psi_{123}^{(0)}(X)\,
\\&&{}\hspace*{-1cm}
\times\Bigl(u_{i\uparrow}^\dagger(1)
u_{j\downarrow}^\dagger(2)d_{k\uparrow}^\dagger(3)-
u_{i\uparrow}^\dagger(1)d_{j\downarrow}^\dagger(2)u_{k\uparrow}^\dagger(3)
\Bigr)\ket{0}\,.
\nonumber
\end{eqnarray}
Here and below the argument of the field $u^\dagger_{i\uparrow}(1)$ etc.
refers to the collection of its arguments that are not shown explicitly,
i.e. $u^\dagger_{\uparrow i}(1)=u^\dagger_{\uparrow i}(x_1,k_{\perp,1})$. The (real) function
$\Psi_{123}^{(0)}(X)$ depends on momentum fractions $x_i$ and transverse momenta $k_{\perp,i}$
of all partons.

Models for $\Psi_{123}^{(0)}(X)$ of various degree of sophistication have been was considered
in different context in a large number of papers see e.g.
Refs.~\cite{Lepage:1980fj,Bolz:1994hb,Bolz:1996sw,Diehl:1998kh,Ji:2003yj,Pasquini:2008ax}.
In this work we adopt the simplest ansatz~\cite{Diehl:1998kh}
\begin{align}\label{Psi0}
\Psi_{123}^{(0)}=&\frac{1}{4\sqrt{6}}\,\phi(x_1,x_2,x_3)\, \Omega_3(a_3,x_i,k_{\perp i})\,.
\end{align}
The transverse momentum dependence is contained in the function
$\Omega_N$
\begin{align}
\Omega_N(a_N,x_i,k_{\perp i})=\frac{(16\pi^2 a_N^2)^{N-1}}{x_1x_2\ldots x_N}
\exp\left[-a_N^2\sum_i k_{\perp i}^2/x_i  \right]\,
\label{eq:Omega}
\end{align}
which is normalized such that
\begin{eqnarray}
&&\int [d^{2}k_{\perp}]_N\Omega_N(a_N,x_i,k_{\perp i})\,=\,1\,,
\nonumber\\
&&\int [d^{2}k_{\perp}]_N\Omega^2_N(a_N,x_i,k_{\perp i})\,=\,\frac{\rho_N}{x_1\ldots x_N}\,,
%\nonumber\\
%&&\hspace*{0.7cm}\rho_N = (8\pi^2 a_N^2)^{N-1},
\label{eq:Omeganorm}
\end{eqnarray}
where $$ \rho_N = (8\pi^2 a_N^2)^{N-1},$$
and $\phi(x_i)$ is related to the leading-twist-3 nucleon distribution
amplitude (see the next Section).
The parameter $a_3$ determines the spread of the wave function in the transverse plane
and e.g. the average quark transverse momentum.

The general classification of Fock states involving an additional gluon
was given in Ref.~\cite{Ji:2003yj}. Unfortunately, we do not agree with the
analysis in~\cite{Ji:2003yj} of the symmetry properties of the corresponding
LCWFs.

As in the three quark case, we restrict ourselves to the states with zero total
orbital angular momentum, $L_z=0$. There are two possibilities~\cite{Ji:2003yj}:
either the quark helicities sum up to $\lambda_{uud}=3/2$
and the gluon has opposite helicity to that of the proton,
$\lambda_g=-1$,  or, alternatively, $\lambda_{uud}= -1/2$ and $\lambda_g=+1$.
We begin with the first case.

The starting observation is that the $SU(3)$ generators obey the following identity
\begin{align}\label{eT}
\epsilon^{ijl} t^a_{lk}+\epsilon^{ilk} t^a_{lj}+\epsilon^{ljk} t^a_{li}=0
\end{align}
As a consequence, there exists only one possibility to form a colorless state
(up to equivalent redefinitions)
%\begin{widetext}
  \begin{eqnarray}\label{gminus}
\ket{p,+}_{uudg_\downarrow} &=&\epsilon^{ijk}\int[\mathcal{D}X]_4
\,\Psi^\downarrow_{1234}(X)\,
\nonumber\\ &&{}\hspace*{-1cm}
\times g_{\downarrow}^{a,\dagger}(4)\,
[t^a u_{\uparrow}(1)]_i^\dagger\, u^\dagger_{j\uparrow}(2) \,d_{k\uparrow}^\dagger(3)\ket{0}\,.
\end{eqnarray}
Note that $[t^a u_{\uparrow}(1)]_i^\dagger = u^\dagger_{i'\uparrow}(1)\,t^a_{i'i}$.
Symmetry properties of the LCWF $\Psi^\downarrow_{1234}$ are determined by the requirement that
the nucleon has isospin 1/2. Since $I_3=1/2$ is fixed by the quark flavor content, 
the $I=1/2$ requirement is equivalent to the simpler condition
that the state is annihilated by the isospin step-up operator
$$
I_+ \ket{p,+}_{uudg_\downarrow} = 0.
$$
 The action of $I_+$ amounts to the replacement
of quark flavors $d\to u$ in~(\ref{gminus}), $I^+\sim u^\dagger {\delta}/{\delta d^\dagger}$.
Projecting the resulting state onto
$\bra{0}g_{\downarrow}^{a'}(4')u_{k'\uparrow}(3')u_{j'\uparrow}(2')u_{k'\uparrow}(1')$
 and collecting the terms in the two independent color structures
 (cf. Eq.~(\ref{eT})) one finds two constraints:
\begin{align}\label{FF}
\Psi^\downarrow_{1234}+\Psi^\downarrow_{1324}-\Psi^\downarrow_{3124}-\Psi^\downarrow_{3214}=0\,,\notag\\
\Psi^\downarrow_{2134}+\Psi^\downarrow_{2314}-\Psi^\downarrow_{3124}-\Psi^\downarrow_{3214}=0\,.
\end{align}
Since the second equation can be obtained from the first one by renaming
$1\leftrightarrow 2$, only one of them is independent.
In order to solve this constraint it is convenient to represent the function
$\Psi^\downarrow$ as sum of contributions with definite parity under cyclic permutations of
the first three (quark) arguments $123\to 231$:
\begin{equation}\label{}
\Psi^\downarrow_{1234}=\Psi^{\downarrow,0}_{1234}+\Psi^{\downarrow,+}_{1234}+\Psi^{\downarrow,-}_{1324}\,,
\end{equation}
such that
$$\Psi^{\downarrow,0}_{1234}=\Psi^{\downarrow,0}_{2314}\,, \qquad
\Psi^{\downarrow,\pm}_{1234}=e^{\pm 2\pi i/3}\Psi^{\downarrow,\pm}_{2314}.$$
One easily finds that an arbitrary function $\Psi^{\downarrow,0}_{1234}$
is a solution of Eq.~(\ref{FF}) whereas
one has to require that $\Psi^{\downarrow,-}_{1234}=-\Psi^{\downarrow,+}_{1324}$.
Thus the most general solution to the isospin constraint can be written as
\begin{equation}\label{eq:Phi}
\Psi^\downarrow_{1234}=\Phi^{\downarrow,0}_{1234}+\Psi^{\downarrow,+}_{1234}-\Psi^{\downarrow,+}_{1324}\,,
\end{equation}
where $\Psi^{\downarrow,0}$ and $\Psi^{\downarrow,+}$ are arbitrary functions with the specified symmetry
under cyclic permutations.

Our result does not agree with the conclusion of~\cite{Ji:2003yj} 
that the function $\Psi^\downarrow_{1234}$
($\psi^{(1)}_{uudg}$ in notations of Ref.~\cite{Ji:2003yj}) is antisymmetric with respect
to permutation of the second and third arguments, which is a much stronger condition.
In fact any function which is antisymmetric in $2\leftrightarrow 3$ can indeed be written in the
form (\ref{eq:Phi}). However, e.g. a totally symmetric function in the quark arguments is
also allowed. The reason why this does not contradict isospin counting is that the corresponding
state is annihilated by $I_+$ thanks to the color identity (\ref{eT}). We note in passing that
 the $SU(3)$-color generators in the definitions given in~\cite{Ji:2003yj} must be transposed,
$t^a_{ii'}\to t^a_{i'i}$.

The second case, a gluon with positive helicity, can be treated similarly.
There exist two independent LCWFs which can be defined as
\begin{widetext}
\begin{eqnarray}\label{gplus}
\ket{p,+}_{uudg^\uparrow}&=&\epsilon^{ijk} \int[\mathcal{D}X]_4\Big\{
\Psi^{\uparrow(1)}_{1234}(X)\,
[t^au_{\downarrow}(1)]_i^\dagger
\Bigl(u_{j\uparrow}^\dagger(2)d_{k\downarrow}^\dagger(3)-d_{j\uparrow}^\dagger(2)u_{k\downarrow}^\dagger(3)
\Bigr)g_{\uparrow}^{a,\dagger}(4)
\notag\\
&&{}\hspace*{10mm}
+\Psi^{\uparrow(2)}_{1234}(X)
u^\dagger_{i\downarrow}(1)\Big( [t^a u_{\downarrow}(2)]^\dagger_j\, d^\dagger_{k\uparrow}(3)-
[t^a d_{\downarrow}(2)]^\dagger_j\, u^\dagger_{k\uparrow}(3)\Big) g_{\uparrow}^{a,\dagger}(4)
\Big\}\ket{0}\,.
\end{eqnarray}
\end{widetext}
The functions $\Psi^{\uparrow(1)}_{1234}$ and $\Psi^{\uparrow(2)}_{1234}$ have no  symmetry constraints.
This result also does not agree with~\cite{Ji:2003yj}.

In what follows we accept the following ansatz for the quark-gluon LCWFs
\begin{align}\label{PhiPsi}
\Psi^\downarrow_{1234}=& \frac1{\sqrt{2x_4}}  \phi_g(x_1,x_2,x_3,x_4)\, \Omega_4(a_g^\downarrow,x_i,k_{\perp i})\,,
\nonumber\\
\Psi^{\uparrow(1)}_{1234}=& \frac1{\sqrt{2x_4}}   \psi_g^{(1)}(x_1,x_2,x_3,x_4)\, \Omega_4(a_g^\uparrow,x_i,k_{\perp i})\,,
\nonumber\\
\Psi^{\uparrow(2)}_{1234}= &  \frac1{\sqrt{2x_4}}  \psi_g^{(2)}(x_1,x_2,x_3,x_4)\, \Omega_4(a_g^\uparrow,x_i,k_{\perp i})\,.
\end{align}
The function $\Omega_4$ is defined in Eq.~(\ref{eq:Omega}) and the momentum fraction distributions
$\phi_g(x_i)$, $\psi_g^{(1,2)}(x_i)$ are related to the next-to-leading twist-4 nucleon distribution
amplitudes as discussed in the next section. For simplicity we choose the same parameter $a_g^\uparrow$
determining the spread of both wave functions $\Psi^{\uparrow(1)}$ and $\Psi^{\uparrow(2)}$ in the transverse
plane. This restriction can be relaxed.

%%%%%%%%%%%%%%%%%%%%%%%%%%%%%%%%%%%%%%%%%%%%%%%%%%%%%%%%%%%%%%%%%%%%%%%%%%%%%%%%%%%%%%%%%%%%%%%
%%%%%%%%%%%%%%%%%%%%%%%%%%%%%%%%%%%%%%%%%%%%%%%%%%%%%%%%%%%%%%%%%%%%%%%%%%%%%%%%%%%%%%%%%%%%%%%
\subsection{Relation to nucleon distribution amplitudes}
%%%%%%%%%%%%%%%%%%%%%%%%%%%%%%%%%%%%%%%%%%%%%%%%%%%%%%%%%%%%%%%%%%%%%%%%%%%%%%%%%%%%%%%%%%%%%%%
%%%%%%%%%%%%%%%%%%%%%%%%%%%%%%%%%%%%%%%%%%%%%%%%%%%%%%%%%%%%%%%%%%%%%%%%%%%%%%%%%%%%%%%%%%%%%%%

Nucleon distribution amplitudes (DAs) are defined as LCWFs with all constituents at small
transverse separations, schematically \cite{Lepage:1980fj}
\begin{equation}
  \phi(x_i,\mu) \sim \int^{|k_\perp|<\mu}[d k_\perp]_N\, \Phi_N(x_i,k_{\perp,i})\,.
\end{equation}
As always in a field theory, taking an asymptotic limit (here vanishing transverse distance)
produces divergences that have to be regularized. Hence DAs are scale-dependent objects
which only include contributions of small transverse momenta,  less that the cutoff.

The exponential ansatz for the the transverse momentum dependence of the LCWFs
(\ref{Psi0}), (\ref{PhiPsi}) implicitly assumes that contributions of hard
gluon exchanges $\sim 1/k_\perp^2$ are subtracted as well,
so that it is natural to identify integrals of the LCWFs over transverse momenta
with the corresponding DAs at a certain low normalization scale.
The advantage of  of imposing this condition is that nucleon DAs allow for a different and
more rigorous definition in terms of matrix elements of nonlocal light-ray operators.
Their moments can be studied using Wilson operator product expansion (OPE) and estimated
using QCD sum rules and/or lattice calculations.
The identification of the integrals of the LCWFs with (dimensionally regularized)
DAs can be viewed as the choice of a specific renormalization (factorization) scheme.

To begin with, consider the leading twist-three nucleon DA which is defined
by the matrix element~\cite{Braun:2000kw}
\begin{eqnarray}
\lefteqn{\vev{0| \epsilon^{ijk}\,
\left(u^{\uparrow,T}_i(z_1 n) C \slashed{n} u^{\downarrow}_j(z_2 n)\right)
\slashed{n} d^{\uparrow}_k(z_3 n) |p}=}
\nonumber\\
&=& - \frac12 p_+ \slashed{n} N^\uparrow(p)\,\int [dx]_3 \,e^{-i p_+ \sum x_i z_i}\,
\Phi_3(x_i)\,,
\end{eqnarray}
where $N(p)$ is the nucleon Dirac spinor, $p^2=m_N^2$, $N^\uparrow(p) = \frac12 (1+\gamma_5)N(p)$ and
$C$ is the charge conjugation matrix. Going over to the two-dimensional spinor notation~(\ref{lambdamu})
and using the explicit expression for the $C$-matrix in Weyl representation~\cite{C-matrix}
\begin{equation}
C=i\gamma^2\gamma^0=\begin{pmatrix}\epsilon_{\alpha\beta}&0\\0&\epsilon^{\dot\alpha\dot\beta}
\end{pmatrix}
\label{eq:Cmatrix}
\end{equation}
this definition can be rewritten equivalently as
\begin{eqnarray}\label{DA3}
\lefteqn{\hspace*{-1cm}
\vev{0|\epsilon^{ijk} u_+^{i\uparrow }(z_1)u_+^{j\downarrow }(z_2)d_+^{k \uparrow }(z_3)|p,+}=}
\nonumber\\
&=&
\frac1{\sqrt{2}} p_+^{3/2} \int [dx]_3\, e^{-ip_+\sum x_i z_i} \Phi_3({x})\,,
\end{eqnarray}
where we suppressed, for brevity, the light-like vector in the arguments of the fields, i.e.
$u_+^{i\uparrow }(z_1)\equiv u_+^{i\uparrow }(n z_1)$, etc.

Making use of~(\ref{qupdown}) and the explicit expression
for the proton state in (\ref{Ansatz}) one finds after a short calculation
\begin{align}\label{}
\phi(x_1,x_2,x_3) =  \Phi_3(x_1,x_2,x_3;\mu_0)\,,
\end{align}
i.e. the function $\phi(x_i)$ which enters the definition
(\ref{Psi0}) of the three-quark LCWF is nothing but the leading-twist nucleon DA.

The DA $\Phi_3(x;\mu)$ can be expanded in eigenfunctions of the one-loop
evolution kernel $P_k(x)$ such that the coefficients $c_k(\mu)$
have autonomous scale dependence:
\begin{equation}\label{phi3q}
\Phi_3(x;\mu)=
120\,  x_1x_2x_3\,\sum_{k=0}^\infty
\left(\frac{\alpha_s(\mu)}{\alpha_s(\mu_0)}\right)^{\gamma_k/\beta_0}
\!\!\!\!\!\!c_k(\mu_0)\,P_k(x)\,.
\end{equation}
The eigenfunctions $P_k(x)$ form a specific set of homogeneous
polynomials of three variables which are orthogonal with respect to the
conformal scalar product~\cite{Braun:1999te}:
\begin{eqnarray}
 120 \int[dx]_3\, x_1 x_2 x_3\, P_k(x) P_j(x) = \nu_k \delta_{kj}\,,
\label{eq:nuk}
\end{eqnarray}
where the coefficients $\nu_k$ depend on the normalization convention for the
eigenfunctions $P_k(x)$. One can show that all eigenfunctions
have definite parity under the interchange of the first and the
third argument: $P_k(x_3,x_2,x_1)= \pm P_k(x_1,x_2,x_3)$.
The first few terms in this expansion are~\cite{Braun:2008ia}
\begin{eqnarray}\label{BK3phi}
\Phi_3(x_1,x_2,x_3)&=& 120 \,f_N\, x_1x_2x_3 \Big[1+a\frac{3}4(x_1-x_3)
\nonumber\\&&{}
+b\frac14(x_1+x_3-2x_2)+\ldots\Big]
\end{eqnarray}
where we have changed the notation to $f_N= c_0$, $a=c_1/c_0$, $b=c_2/c_0$.
The corresponding anomalous dimensions are
$\gamma_{0}=2/3$, $\gamma_a=20/9$ and $\gamma_b=8/3$.

The normalization constant $f_N$ is determined by the matrix element of the
corresponding local three-quark operator. It was calculated several times
in the past using QCD sum rules
~\cite{Chernyak:1984bm,King:1986wi,Chernyak:1987nu,Braun:2006hz,Gruber:2010bj}:
\begin{equation}
 f_N  =  \int [dx]_3 \Phi_3(x) = (5.0\pm 0.5)\times 10^{-3}\,\, \text{GeV}^2.
\label{eq:fN}
\end{equation}
The latest estimates for the ``shape'' parameters $a, b$ from lattice calculations
\cite{Braun:2008ur,Braun:2010hy} are in the range
\begin{equation}
   \frac34 a \,=\, 0.85-0.95\,, \qquad \frac14 b \,=\, 0.23-0.33\,.
\label{eq:ab}
\end{equation}
These values are consistent with the light-cone sum rules for nucleon electromagnetic
form factors~\cite{Braun:2006hz} and somewhat smaller than the earlier QCD sum rule
estimates~\cite{Chernyak:1984bm,King:1986wi,Chernyak:1987nu}.

The model used in Ref.~\cite{Diehl:1998kh,Bolz:1996sw} corresponds to
$a = b = 1$ at the scale $\mu_0=1\text{GeV}$ which does not contradict (\ref{eq:ab}).
The overall normalization constant was determined in Ref.~\cite{Diehl:1998kh}
from the fit to parton distributions at large values of Bjorken $x$:
$f_N=4.7\times 10^{-3}\,\, \text{GeV}^2$,
in a remarkably good agreement with Eq.~(\ref{eq:fN}).
This agreement is very encouraging as a strong indication for
the selfconsistence of the whole approach.
Note that the coupling $f_N$ is related to the normalization constant $f_3$ used 
in~\cite{Bolz:1996sw,Diehl:1998kh} as $f_3 = \sqrt{2}f_N$.

The quark-gluon twist-4 nucleon DAs were introduced
in~\cite{Braun:2008ia}
\begin{widetext}
\begin{align}
\vev{0|ig\epsilon^{ijk}u^{\downarrow i}_{+}(z_1)\,u^{\uparrow j}_{+}(z_2)\,
[\bar f_{++}(z_4)\, {d}^{\downarrow}_{+}(z_3)]^k|p,\lambda}=&
-\frac14
m_N p_+^2 N^{\uparrow}_+(p)\int [dx]_4 \, e^{-ip_+\,\sum {x_iz_i}}\,\Phi^g_4(x)\,,
\nonumber\\
\vev{0|ig\epsilon^{ijk}\,u^{\uparrow i}_{+}(z_1)\, [\bar f_{++}(z_4)u^{\downarrow }_{+}(z_2)]^j\,
{d}^{\downarrow k}_{+}(z_3) \,|p,\lambda}=&
-\frac14
m_N p_+^2 N^{\uparrow}_+(p)\int [dx]_4 \, e^{-ip_+\,\sum {x_iz_i}}\,\Psi^g_4(x)\,,
\nonumber\\
\vev{0|ig\epsilon^{ijk} [\bar f_{++}(z_4) \,u^{\downarrow}_+(z_1)]^i\, u^{\downarrow j}_+(z_2)\,
{d}^{\downarrow k}_{+}(z_3)|p,\lambda}=
&
-\frac14
m_N p_+^2 N^{\downarrow}_+(p)\int [dx]_4 \, e^{-ip_+\,\sum {x_iz_i}}\,\Xi^g_4(x)\,,
\label{BMRa}
\end{align}
\end{widetext}
where we changed an overall sign because of the different definition of the
charge conjugation matrix, cf.~\cite{C-matrix}.

The asymptotic DAs are
\begin{align}\label{asDAs}
\Phi_4^g(x,\mu) &= - \frac14 8!\, x_1 x_2 x_3 x_4^2\, \Big[\lambda_2^g(\mu) - \frac13 \lambda_3^g(\mu)\Big]\,,
\nonumber\\
\Psi_4^g(x,\mu) &= \phantom{+} \frac14 8!\, x_1 x_2 x_3 x_4^2\, \Big[\lambda_2^g(\mu) + \frac13 \lambda_3^g(\mu)\Big]\,,
\nonumber\\
\Xi_4^g(x,\mu)  &= \phantom{+} \frac16 8!\, x_1 x_2 x_3 x_4^2 \, \lambda_1^g(\mu) \,,
\end{align}
where $\lambda^g_k$ are multiplicatively renormalizable
couplings
\begin{align}
  \lambda_{1}^g(\mu) &= \lambda_{1}^g(\mu_0)\, L^{19/(3\beta_0)}\,,
\nonumber\\
  \lambda_{2}^g(\mu) &= \lambda_{2}^g(\mu_0)\, L^{7/\beta_0}\,,
\nonumber\\
  \lambda_{3}^g(\mu) &= \lambda_{3}^g(\mu_0)\, L^{79/(9\beta_0)}\,,
\label{eq:gluon_adim}
\end{align}
with $L = \alpha_s(\mu)/\alpha_s(\mu_0)$ and  $\beta_0 = 11 - \frac23 n_f$.
In notation of Ref.~\cite{Braun:2008ia}
$\lambda_1^g = \xi_{2}^g$, $\lambda_2^g = \eta_{2,0}^g$ and  $\lambda_3^g = \eta_{2,1}^g$.

Numerical values of these parameters can be estimated using QCD sum rules, see App.~A.
We obtain at the scale $\sim 1$~GeV:
\begin{align}
    \lambda_1^g = &  (2.6\pm 1.2)\cdot 10^{-3}\text{GeV}^2\,,
\notag\\
    \lambda_2^g = &  (2.3\pm 0.7)\cdot 10^{-3}\,\text{GeV}^2\,,
\notag\\
    \lambda_3^g = &  (0.54\pm  0.2)\cdot 10^{-3}\,\text{GeV}^2\,,
\label{eq:lambdag}
\end{align}
where the sign convention is that the three-quark coupling $f_N$ is positive.

Evaluating the matrix elements in the definitions of DAs~(\ref{BMRa})
using (\ref{qupdown}),~(\ref{a+a}) and explicit expressions for the $uudg$ Fock states
in terms of the corresponding LCWFs one obtains the required relations:
\begin{eqnarray}\label{psiPsi}
\lefteqn{{g}\,\phi_g(x_1,x_3,x_2,x_4)=}
\nonumber\\&=&
-\frac{m_N}{96}\Big[2\Xi_4^g(x_1,x_2,x_3,x_4) + \Xi_4^g(x_2,x_1,x_3,x_4)\Big]\,,
\notag\\
\lefteqn{{g}\,\psi_g^{(1)}(x_1,x_2,x_3,x_4)=}
\nonumber\\&=&
-\frac{m_N}{48}\Big[\Psi_4^g(x_2,x_1,x_3,x_4)+\frac12\Phi_4^g(x_1,x_2,x_3,x_4) \Big]\,,
\notag\\
\lefteqn{{g}\,\psi_g^{(2)}(x_1,x_3,x_2,x_4)=}
\nonumber\\&=&
\frac{m_N}{48}\Big[\Phi_4^g(x_1,x_2,x_3,x_4)+\frac12\Psi_4^g(x_2,x_1,x_3,x_4)\Big]\,.
\end{eqnarray}
Note that the DA $\Xi_4^g$ satisfies the symmetry relation~\cite{Braun:2008ia}
\begin{eqnarray}
\lefteqn{\hspace*{-1cm}\Xi_4^g(x_1,x_2,x_3,x_4)+\Xi_4^g(x_1,x_3,x_2,x_4)=}
\nonumber\\&= & \Xi_4^g(x_2,x_3,x_1,x_4)+\Xi_4^g(x_3,x_2,x_1,x_4)\,
\end{eqnarray}
which is consistent with~Eq.~(\ref{FF}).

%%%%%%%%%%%%%%%%%%%%%%%%%%%%%%%%%%%%%%%%%%%%%%%%%%%%%%%%%%%%%%%%%%%%%%%%%%%%%%%%%%%%%%%%%%%%%%%
%%%%%%%%%%%%%%%%%%%%%%%%%%%%%%%%%%%%%%%%%%%%%%%%%%%%%%%%%%%%%%%%%%%%%%%%%%%%%%%%%%%%%%%%%%%%%%%
\subsection{Fock state probabilities}
%%%%%%%%%%%%%%%%%%%%%%%%%%%%%%%%%%%%%%%%%%%%%%%%%%%%%%%%%%%%%%%%%%%%%%%%%%%%%%%%%%%%%%%%%%%%%%%
%%%%%%%%%%%%%%%%%%%%%%%%%%%%%%%%%%%%%%%%%%%%%%%%%%%%%%%%%%%%%%%%%%%%%%%%%%%%%%%%%%%%%%%%%%%%%%%

Our conventions correspond to the usual relativistic normalization of the proton state
\begin{equation}
\vev{p,+|p',+}=(2\pi)^3\,2p_+\delta(p_+-p'_+)\,\delta^2(\vec{p}_\perp-\vec{p'}_\perp)\,.
\end{equation}
The partial contribution of each Fock state is defined similarly, e.g.
\begin{multline}
\vev{p,+|p',+}_{uud}\,=
\\
=\,(2\pi)^3\,2p_+\delta(p_+-p'_+)\,\delta^2(\vec{p}_\perp-\vec{p'}_\perp)\, P_{uud}
 \,,
\end{multline}
where $P_{uud}$
is the probability of the three-quark state with zero orbital angular
momentum.

Using the definition in Eq.~(\ref{Ansatz}) and the ansatz in Eq.~(\ref{Psi0})
we get after the integration over transverse momenta
\begin{align}\label{norm3}
P_{uud}
=& \frac1{96} \rho_3\,f_N^2
\int \frac{[dx]_3}{x_1 x_2 x_3}\Big[ |\phi(x_1,x_2,x_3)|^2
\nonumber\\
&+
\frac12|\phi(x_3,x_2,x_1)+ \phi(x_1,x_2,x_3)|^2\Big],
\end{align}
where $\rho_3 \equiv \rho_{N=3}$ is defined in Eq.~(\ref{eq:Omeganorm}).

{}For the model specified in Eq.~(\ref{phi3q}) one obtains
\begin{align}
P_{uud}
=
\frac{15}{4}f_N^2\rho_3 \left(1+\frac{a^2+b^2}{56}\right).
\end{align}
{}For a given value of the wave function at the origin, $f_N$, the probability
of the three-quark valence state is proportional to the fourth power of the $a_3$ parameter,
$P_{uud}\sim a_3^4$.
We fix $a_3$ from the requirement to have the same probability of
three-quark state as in~\cite{Bolz:1996sw,Diehl:1998kh}. Namely,
for $f_N = 5\times 10^{-3}$~GeV$^2$ and  $a = b = 1$ one gets
\begin{equation}
P_{uud}
=\frac{435}{112}f_N^2\rho_3 \simeq 0.17
\label{eq:Puud}
\end{equation}
for
\begin{equation}
a_3=0.73~\text{GeV}^{-1}.
\label{eq:a3}
\end{equation}

The dependence on the shape of the DA (for $a,b\sim 1$) is very weak.
This property is due to an attractive feature of the
Bolz-Kroll ansatz~(\ref{Psi0}): Different terms in the expansion of the DA in
multiplicatively renormalizable operators (\ref{phi3q}) contribute to the norm additively;
there is no interference.
For the general case one obtains
\begin{align}\label{p3l0}
P_{uud}=\frac54\rho_3\sum_k\Big(3\nu_k^+ |c^+_{k}(\mu)|^2 +\nu_k^- |c^-_{k}(\mu)|^2 \Big )\,,
\end{align}
where $c_k^\pm$ are the expansion coefficients corresponding
to the eigenfunctions $P^\pm_k$  with positive (negative)
parity with respect to the permutation $x_1\leftrightarrow x_3$:
Each state with positive parity contributes with extra factor three.

The probabilities of the four-parton states with an extra gluon with negative (positive) helicity
are given by
\begin{eqnarray}
P_{uudg_\downarrow}&=&
   2\rho_4\int \frac{[dx]_4}{w(x)}\phi_g(x)\big(2-\mathcal{P}_{12}\big) \phi_g(x)\,,
\nonumber\\
P_{uudg^\uparrow}&=&
   4\rho_4 \int \frac{[dx]_4}{w(x)}\Big\{\Big(\psi_g^{(1)}+\mathcal{P}_{23}\psi_g^{(2)}(x)\Big)^2
\nonumber\\&&
{}\hspace*{-5mm}-\psi_g^{(2)}(x)\mathcal{P}_{23}\psi_g^{(1)}(x)
  +\psi_g(x)\Big(1\!-\!\frac12\mathcal{P}_{13}\Big)\psi_g(x)
\Big\},
\nonumber\\
\label{PPg}
\end{eqnarray}
where we use a shorthand notation
\begin{equation}
w(x)=x_1x_2x_3x_4^2\,
\end{equation}
and
\begin{align}\label{def:varphi}
\psi_g(x)=\psi^{(1)}_g(x_1,x_2,x_3,x_4)-\psi^{(2)}_g(x_3,x_1,x_2,x_4)\,.
\end{align}
Here and below $\mathcal{P}_{12}$, $\mathcal{P}_{23}$, etc. are quark permutation operators, e.g.
$\mathcal{P}_{12}\,\psi_g(x_1,x_2,x_3,x_4) = \psi_g(x_2,x_1,x_3,x_4)$.
We also assumed  that the functions $\phi_g$ and $\psi_g^{(1,2)}$ are real.

With the help of Eqs.~(\ref{psiPsi}) one can rewrite these expressions in terms
of the nucleon DAs,  $\Xi_4^g$, $\Psi_4^g$ and $\Phi_4^g$.
{}Using asymptotic DAs specified in~(\ref{asDAs}) and the value $\alpha_s = 0.5$ (at the scale 1 GeV)
one obtains for the central values of the couplings in Eq.~(\ref{eq:lambdag})
\begin{eqnarray}
P_{uudg^\downarrow} &\!=\!& \frac{35}{8g^2}m_N^2\rho_4 (\lambda^g_1)^2
\simeq 0.30 \left(\frac{a^\downarrow_g}{a_3}%{0.75~\text{\small GeV}^{-1}}
\right)^6\!\!,
\nonumber\\
P_{uudg^\uparrow}  &\!=\!& \frac{105}{16g^2}m_N^2\rho_4
\Big[(\lambda^g_2)^2 \!+\! (\lambda^g_3)^2\Big]
\simeq 0.37 \left(\frac{a^\uparrow_g}%{0.75~\text{\small GeV}^{-1}}
{a_3}\right)^6\!\!\!,
\nonumber\\
\end{eqnarray}
where $a_3=0.73~\text{\small GeV}^{-1}$, Eq.~(\ref{eq:a3}).
The choice $a_g^\downarrow = a_g^\uparrow = a_3$
corresponds to the same spread in transverse plane as for the three-quark wave function.
These numbers are of the right order of magnitude, which is encouraging.

For the general case, the DAs $\Xi_4^g$, $\Psi_4^g$ and $\Phi_4^g$
can be expanded in contributions of multiplicatively renormalizable operators
as follows~\cite{Braun:2008ia}
\begin{eqnarray}\label{XI}
\Xi_4^g(x_1,x_2,x_3,x_4;\mu )&=& \phi_0(x)\sum_k c^\Xi_k(\mu)\, P^\Xi_k(x)\,,
\nonumber\\
\Psi_4^g(x_1,x_2,x_3,x_4;\mu)&\pm& \Phi_4^g(x_3,x_1,x_2,x_4;\mu)
\nonumber\\
&=&\phi_0(x)\sum_{k} c^\pm_k(\mu)\, P^\pm_k(x)\,,
\end{eqnarray}
where $\phi_0(x) = \frac12 8! x_1 x_2 x_3 x_4^2$ and $P_k^{\Xi,\pm}(x_1,x_2,x_3,x_4)$
are orthogonal polynomials which we assume normalized as
\begin{eqnarray}
\nu^\Xi_k \delta_{kr}&=& \int[dx]_4\,\phi_0(x)\,P^\Xi_k(x)\,(2+\mathcal{P}_{12})\,P^\Xi_r(x)\,,
\nonumber\\
\nu^\pm_k \delta_{kr}&=& \int[dx]_4\,\phi_0(x)\,P_k^\pm(x)\,(2\pm \mathcal{P}_{23})\,P_r^\pm(x)\,.
\end{eqnarray}

Inserting~(\ref{psiPsi}) and~(\ref{XI}) into~(\ref{PPg}) one finds after some algebra
\begin{align}\label{}
P_{uudg^\downarrow}=&\frac{105m_N^2\rho_4}{8g^2}\sum_{k} \nu^\Xi_k |c^\Xi_k|^2\,,
\\
P_{uudg^\uparrow}=& \frac{105m_N^2\rho_4}{4 g^2}\biggl[\sum_{k}3\nu^+_k|c^+_k|^2 +\sum_{k} \nu^-_r|c^-_k|^2\biggr].
\nonumber
\end{align}
Similar to the three-quark case, each multiplicatively renormalizable contribution to the DA
generates an additive contribution to the state probability; there is no interference.

%%%%%%%%%%%%%%%%%%%%%%%%%%%%%%%%%%%%%%%%%%%%%%%%%%%%%%%%%%%%%%%%%%%%%%%%%%%%%%%%%%%%%%%%%%%%%%%
%%%%%%%%%%%%%%%%%%%%%%%%%%%%%%%%%%%%%%%%%%%%%%%%%%%%%%%%%%%%%%%%%%%%%%%%%%%%%%%%%%%%%%%%%%%%%%%
\section{Parton densities}
%%%%%%%%%%%%%%%%%%%%%%%%%%%%%%%%%%%%%%%%%%%%%%%%%%%%%%%%%%%%%%%%%%%%%%%%%%%%%%%%%%%%%%%%%%%%%%%
%%%%%%%%%%%%%%%%%%%%%%%%%%%%%%%%%%%%%%%%%%%%%%%%%%%%%%%%%%%%%%%%%%%%%%%%%%%%%%%%%%%%%%%%%%%%%%%

The definitions of quark and gluon parton densities can be found e.g. in the review
\cite{Diehl:2003ny}. Translating them into the two-component spinor notation
we obtain for quark and gluon distributions
\begin{widetext}
\begin{eqnarray}
 q(x) &=&
\frac{1}{2} \int \frac{d z}{2\pi} e^{ixz(pn)}
\langle p| \bar q_+^\uparrow(-\frac12 zn)q_+^\uparrow(\frac12 zn) + \bar q_+^\downarrow(-\frac12 zn)q_+^\downarrow(\frac12 zn)
|p\rangle\,,
\nonumber \\
 \Delta q(x) &=&
\frac{1}{2} \int \frac{d z}{2\pi} e^{ixz(pn)}
\langle p,+| \bar q_+^\uparrow(-\frac12 zn)q_+^\uparrow(\frac12 zn) - \bar q_+^\downarrow(-\frac12 zn)q_+^\downarrow(\frac12 zn)
|p,+\rangle\,,
\nonumber \\
 \delta q(x) &=&
\frac{1}{2} \int \frac{d z}{2\pi} e^{ixz(pn)}
\langle p,+| \bar q_+^\uparrow(-\frac12 zn)q_+^\downarrow(\frac12 zn)|p,-\rangle\,,
\end{eqnarray}
\begin{eqnarray}
  x g(x) &=& \frac{1}{2pn} \int \frac{d z}{2\pi} e^{ixz(pn)}
\langle p|f^a_{++}(-\frac12 zn) \bar f^a_{++}(\frac12 zn) + \bar f^a_{++}(-\frac12 zn) f^a_{++}(\frac12 zn)|p\rangle\,,
\nonumber\\
  x \Delta g(x) &=& \frac{1}{2pn} \int \frac{d z}{2\pi} e^{ixz(pn)}
\langle p,+|f^a_{++}(-\frac12 zn) \bar f^a_{++}(\frac12 zn) - \bar f^a_{++}(-\frac12 zn) f^a_{++}(\frac12 zn)|p,+\rangle\,,
\end{eqnarray}
\end{widetext}
respectively. Here $q(x)$, $g(x)$ are unpolarized and $\Delta q(x)$, $\Delta g(x)$ polarized densities,
and $\delta q(x)$ is the quark transversity. For the unpolarized distributions the average over the
proton polarizations is assumed.

The quark parton distributions for each flavor $q=u,d$ receive contributions
from the three-quark $3q \equiv uud$ Fock state and also from the $3qg \equiv uudg$
states with both gluon helicities:
\begin{eqnarray}
q(x) &=& q_{3q}(x) + q_{3qg_\downarrow}(x) +  q_{3qg_\uparrow}(x)\,,
\end{eqnarray}
and similar for $\Delta q(x)$ and  $\delta q(x)$. The three-quark contributions are:
\begin{widetext}
\begin{eqnarray}
\begin{pmatrix} u_{3q}(x) \\ \Delta u_{3q}(x)\end{pmatrix}&=&
\frac{\rho_3 f_N^2}{96x}\int_0^{1} \frac{dx_2 dx_3}{x_2 x_3} \delta(1-x-x_2-x_3)\Big\{\phi^2(x,x_2,x_3)
\pm\phi^2(x_2,x,x_3)+\Big[\phi(x,x_3,x_2)+\phi(x_2,x_3,x)\Big]^2\Big\},
\nonumber\\
\begin{pmatrix} d_{3q}(x) \\ \Delta d_{3q}(x)\end{pmatrix}&=&
\frac{\rho_3 f_N^2}{96x}\int_0^{1} \frac{dx_1 dx_2}{x_1 x_2} \delta(1-x-x_1-x_2)
\Big\{\phi^2(x_1,x_2,x)
\pm\frac12\Big[\phi(x_1,x,x_2)+\phi(x_2,x,x_1)\Big]^2 \Big\},
\end{eqnarray}
and
\begin{eqnarray}
\delta u_{3q}(x)&=&\frac{\rho_3 f_N^2}{96x}\int_0^{1} \frac{dx_2 dx_3}{x_2 x_3} \delta(1-x-x_2-x_3)
\Big[\phi(x,x_2,x_3)+\phi(x_3,x_2,x)\Big]\Big[\phi(x,x_3,x_2)+\phi(x_2,x_3,x)\Big]\,,
\notag\\
\delta d_{3q}(x)&=&-\frac{\rho_3 f_N^2}{96x}\int_0^{1} \frac{dx_1 dx_2}{x_1 x_2} \delta(1-x-x_1-x_2)
\phi(x_1,x_2,x)\phi(x_2,x_1,x)\,.
\end{eqnarray}
For the three-quark-gluon contributions we obtain:
\begin{eqnarray}
\begin{pmatrix} u_{3qg_\downarrow}(x) \\ \Delta u_{3qg_\downarrow}(x)\end{pmatrix}&=&
2\rho_4\int \frac{[dx]_4}{w(x)}[\delta(x-x_1)+\delta(x-x_2)]
\phi_g(x)\left(2-\mathcal{P}_{12}\right)\phi_g(x)\,,
\nonumber\\
\begin{pmatrix} d_{3qg_\downarrow}(x) \\ \Delta d_{3qg_\downarrow}(x)\end{pmatrix}&=&
2\rho_4\int \frac{[dx]_4}{w(x)}\delta(x-x_3)
\phi_g(x)\left(2-\mathcal{P}_{12}\right)\phi_g(x)\,,
\nonumber\\
\begin{pmatrix} u_{3qg_\uparrow}(x) \\ \Delta u_{3qg_\uparrow}(x)\end{pmatrix}&=&
2\rho_4\int \frac{[dx]_4}{w(x)}\Big\{2\big[\delta(x-x_2)\pm\delta(x-x_1)\big]
\Big[\Big(\psi_g^{(1)}(x)+\mathcal{P}_{23}\psi_g^{(2)}(x)\Big)^2-
\psi_g^{(1)}(x) \mathcal{P}_{23}\psi_g^{(2)}(x)\Big]
\nonumber\\&&{}\hspace*{1.8cm}
\pm\big[\delta(x-x_1)+\delta(x-x_3)\big]
\psi_g(x)\left(2-\mathcal{P}_{13}\right)\psi_g(x)\Big\},
\nonumber\\
\begin{pmatrix} d_{3qg_\uparrow}(x) \\ \Delta d_{3qg_\uparrow}(x)\end{pmatrix}&=&
2\rho_4\int\frac{[dx]_4}{w(x)}\Big\{\delta(x-x_2)\,\psi_g(x)\left(2-\mathcal{P}_{13}\right)\psi_g(x)
\pm 2\delta(x-x_3)\Big[\Big(\psi_g^{(1)}(x)+\mathcal{P}_{23}\psi_g^{(2)}(x)\Big)^2
\nonumber\\&&{}\hspace*{1.8cm}
- \psi_g^{(1)}(x) \mathcal{P}_{23}\psi_g^{(2)}(x)\Big]\Big\},
\end{eqnarray}
and
\begin{eqnarray}
\delta d_{3qg^\uparrow}(x)=\delta d_{3qg^\downarrow}(x)&=&
-2\rho_4\int\frac{[dx]_4}{w(x)}\delta(x-x_2)
\Big(\mathcal{P}_{23}\phi_g(x)\Big)\Big(2-\mathcal{P}_{13}\Big)
\psi_g(x)\,,
\notag\\
\delta u_{3qg^\uparrow}(x)=\delta u_{3qg^\downarrow}(x)&=&
2\rho_4\int\frac{[dx]_4}{w(x)}\delta(x-x_2)
\left[\psi_g^{(1)}(x)\Big(2-\mathcal{P}_{12}\Big)\phi_g(x)+
\left(\mathcal{P}_{23}\psi_g^{(2)}(x)\right)\Big(1+\mathcal{P}_{12}\Big)\phi_g(x)\right].
\end{eqnarray}
Finally, for the gluon parton distributions we get
\begin{eqnarray}
\begin{pmatrix} g(x) \\ \Delta g(x)\end{pmatrix}&=&
4\rho_4\int \frac{[dx]_4\delta(x-x_4)}{w(x)}
 \Big\{
\Big(\psi_g^{(1)}(x)+\mathcal{P}_{23}\psi_g^{(2)}(x)\Big)^2
-\psi_g^{(2)}(x) \mathcal{P}_{23}\psi_g^{(1)}(x)
+\frac12\psi_g(x)\big(2-\mathcal{P}_{13}\big)\psi_g(x)
\nonumber\\&&{}\hspace*{2.8cm}
\pm \frac12\Big[ \phi_g(x)\big(2-\mathcal{P}_{12}\big)\phi_g(x) \Big] \Big\}.
\end{eqnarray}
\end{widetext}
{}For simple models of the wave functions the integrations over parton momentum fractions
can be carried out explicitly.
In particular using the three-quark wave function from Ref.~\cite{Bolz:1996sw,Diehl:1998kh}
which corresponds to the choice $a=b=1$ in the nucleon DA (\ref{BK3phi}), one obtains
\begin{eqnarray}
u_{3q}(x)&\!=\!&P_{3q}\frac{1960}{29}\, x(1\!-\!x)^3\Big\{
1\!-\!\frac67(1\!-\!x)\!+\!\frac{12}{35}(1\!-\!x)^2\Big\},
\notag\\
d_{3q}(x)&\!=\!&P_{3q}\frac{140}{29}\, x(1\!-\!x)^3
\Big\{1+ 3(1\!-\!x)+\frac{12}5(1\!-\!x)^2 \Big\},
\notag\\
\Delta u_{3q}(x)&\!=\!&{}\!P_{3q}\frac{5600}{87} x(1\!-\!x)^3
\!\Big\{1\!-\!\frac{21}{20}(1\!-\!x)\!+\!\frac9{40}(1\!-\!x)^2\!\Big\},
\notag\\
\Delta d_{3q}(x)&\!=\!&-P_{3q}\frac{140}{87}\, x(1\!-\!x)^3
\Big\{1\!+\!3(1\!-\!x)\!+\!\frac95(1\!-\!x)^2\Big\},
\notag\\
\delta u_{3q}(x)\!&=\!&P_{3q}\frac{3500}{87}x(1\!-\!x)^3\Big\{1\!-\!\frac35(1\!-\!x)\!+\!\frac9{5^3}
(1\!-\!x)^2\Big\},
\notag\\
\delta d_{3q}(x)&\!=\!&-P_{3q}\frac{140}{87}\, x(1\!-\!x)^3
\Big\{1\!+\!3(1\!-\!x)\!+\!\frac95(1\!-\!x)^2\Big\}.
\notag\\
\label{eq:q}
\end{eqnarray}
These expressions coincide with the corresponding ones in Ref.~\cite{Bolz:1996sw,Diehl:1998kh}.
{}For the three-quark-gluon contributions, taking into account Eqs.~(\ref{psiPsi})
and using asymptotic DAs~(\ref{asDAs}) we arrive at
\begin{eqnarray}
   d_{3qg_\downarrow}(x)&=&\frac12 u_{3qg_\downarrow}(x)= 56 P_{3qg_\downarrow} x(1\!-\!x)^6\,,
\nonumber\\
   d_{3qg^\uparrow}(x)&=&\frac12 u_{3qg^\uparrow}(x)= 56 P_{3qg^\uparrow} x(1\!-\!x)^6\,,
\label{eq:Deltaq1}
\end{eqnarray}
\begin{eqnarray}
\Delta d_{3qg_\downarrow}(x)&=&\frac12 \Delta u_{3qg_\downarrow}(x)= d_{3qg_\downarrow}(x)\,,
\nonumber\\
\Delta d_{3qg^\uparrow}(x)&=& -\left(1-\frac43\beta \right)d_{3qg^\uparrow}(x),
\nonumber\\
\Delta u_{3qg^\uparrow}(x)&=& \frac23\, \beta\, u_{3qg^\uparrow}(x)
\label{eq:Deltaq2}
\end{eqnarray}
\begin{eqnarray}
\delta u_{3qg^\uparrow}(x) &=& \delta u_{3qg^\downarrow}(x)
\nonumber\\
&&\hspace*{-1.5cm}=\,84\sqrt{\frac23}
\Big(\sqrt{1-\beta}-1/3\sqrt{\beta}\Big)\sqrt{P_{3qg_\downarrow}P_{3qg_\uparrow}} \, x(1-x)^6\,,
\nonumber\\
\delta d_{3qg^\uparrow}(x) &=& \delta d_{3qg^\downarrow}(x)
\nonumber\\
&&\hspace*{-1.5cm}\,=\,-56\sqrt{\frac23}\sqrt{\beta}\sqrt{P_{3qg_\downarrow}P_{3qg_\uparrow}}\, x(1-x)^6\,,
\end{eqnarray}
and
\begin{eqnarray}
x g(x)&=& 168\,(P_{3qg^\uparrow} + P_{3qg_\downarrow})\,x^3(1\!-\!x)^5\,,
\nonumber\\
x \Delta g(x)&=& 168\,(P_{3qg^\uparrow} - P_{3qg_\downarrow})\,x^3(1\!-\!x)^5\,.
\label{eq:g}
\end{eqnarray}
where we used the notation
\begin{equation}
\beta =  \frac{(\lambda_3^g)^2}{(\lambda_2^g)^2+ (\lambda_3^g)^2} = 0.052\pm 0.030\,.
\label{eq:beta}
\end{equation}
It is easy to check that the Soffer inequality \cite{Soffer:1994ww,Artru:2008cp}
\begin{equation}
q(x) +\Delta q(x) \ge 2|\delta q(x)|
\label{eq:Soffer}
\end{equation}
is fulfilled for arbitrary values of the parameters.

Note that our result for the the large $x$ behavior of quark parton distributions
due to contribution of the quark-gluon Fock states differs from that
in~\cite{Diehl:1998kh}: $(1\!-\!x)^6$ vs. $(1\!-\!x)^7$.

\begin{figure*}[t]
\includegraphics[width=14.9cm,clip=true]{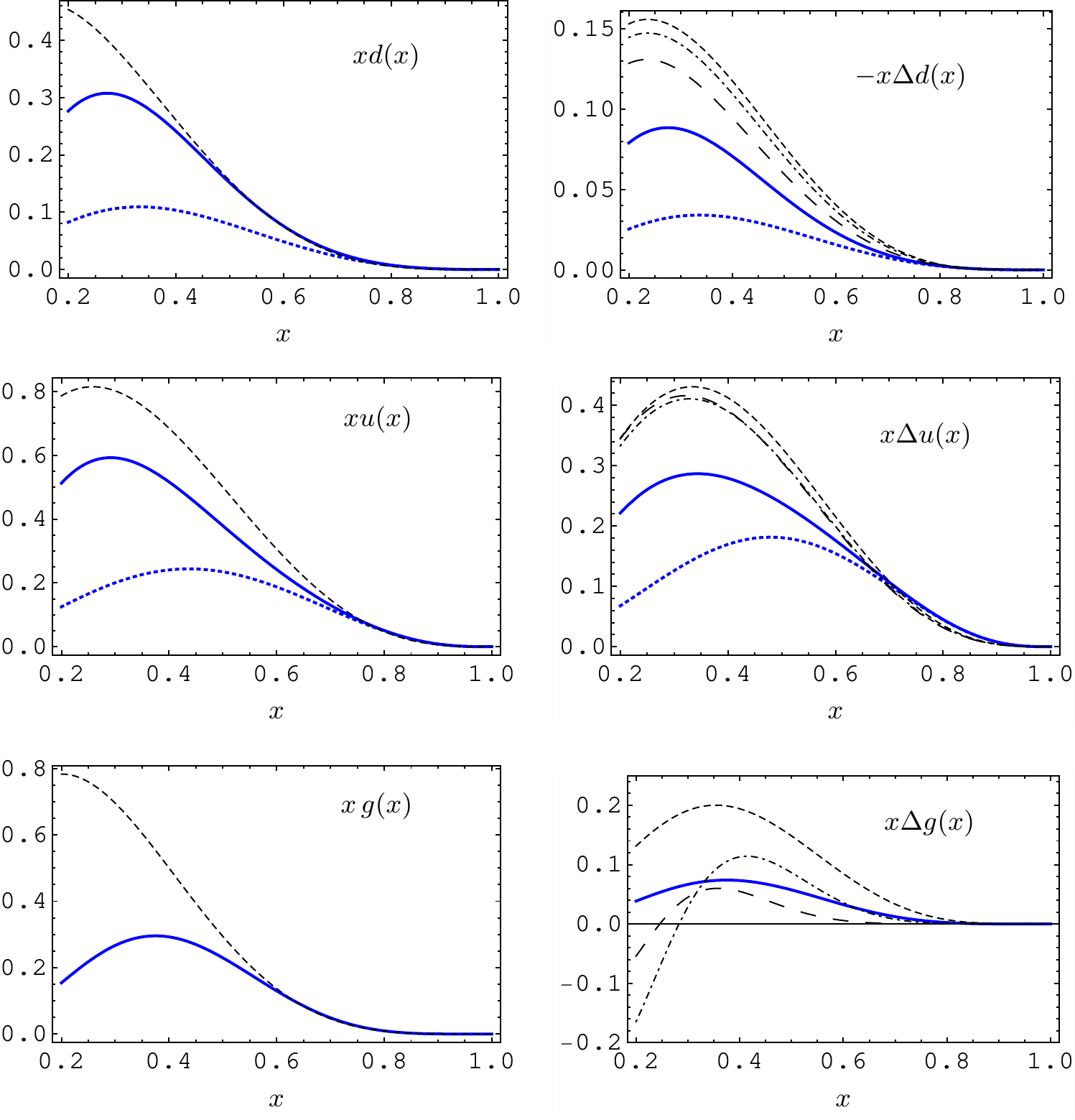}
\caption{Quark and gluon parton distributions. The black
curves correspond to the existing parametrizations: GRV~\cite{Gluck:1998xa} (short dashes),
DSSV~\cite{deFlorian:2009vb} (long dashes) and LSS'10~\cite{Leader:2010rb} (dash-dotted)
at the scale $\mu^2=1GeV^2$~\cite{Cafarella:2003jr}.
The solid blue curve is our model prediction taking into account the
contributions of the valence three-quark state and the state involving one additional gluon.
The contribution of the valence state alone is shown by dots for comparison.
}
\label{fig:PD}
\end{figure*}

{}For the numerical analysis we accept the same three-quark wave function as in
Refs.~\cite{Bolz:1996sw,Diehl:1998kh}, corresponding to the probability of the valence
state $P_{3q}=0.17$ (\ref{eq:Puud}), and fix the remaining parameters of the the
quark-gluon wave functions from the requirement that the resulting parton distributions
are in reasonable agreement with the existing parameterizations at large $x$, see
Fig.~\ref{fig:PD}.
%%%%%%%%%%%%%%%%%%%%%%%%%%%%%%%%%%%%%%%%%%%%%%%%%%%%%%%%%%%%%%%%%%%%%%%%%
%%%%%%%%%%%%%%%%%%%%%%%%%%%%%%%%%%%%%%%%%%%%%%%%%%%%%%%%%%%%%%%%%%%%%%%%%%%%%

The unpolarized distributions are only sensitive to the
total probability to find an extra gluon. We choose
\begin{equation}
 P_{3qg} = P_{3qg^\uparrow}+ P_{3qg^\downarrow} = 0.33\,.
\end{equation}
For the central values of the QCD sum rule
estimates for the wave functions at the origin, Eq.~(\ref{eq:lambdag}),
this value can be obtained assuming that the
quark-gluon state is slightly more compact in transverse space
as compared to the valence three-quark configuration:
\begin{equation}
 a_g = a_g^\uparrow = a_g^\downarrow = 0.9\, a_3\,,
\label{eq:ag}
\end{equation}
which is reasonable.

The ratio $\Delta g(x)/g(x)$ is determined in our simple model
by the ratio of the probabilities to find a gluon with helicity
aligned an anti-aligned with that of the proton.
In the rest of this work we take
\begin{equation}
\frac{P_{uudg^\downarrow}}{P_{uudg^\uparrow}} = \frac23
\frac{(\lambda_1^g)^2}{(\lambda_2^g)^2+ (\lambda_3^g)^2} = 0.6\,\, (0.8\pm 0.2)
\end{equation}
where the number in parenthesis is the QCD sum rule prediction,
Eq.~(\ref{eq:lambdag}).
The polarized quark distributions
$\Delta u(x)$ and $\Delta d(x)$ also involve another ratio of the couplings,
cf.~Eq.~(\ref{eq:beta}), which is, however, small according to our estimates.
The corresponding contributions to $\Delta u(x)$ and $\Delta d(x)$ are below 5\%.

\begin{figure*}[t]
\begin{center}
\includegraphics[width=14cm,clip=true]{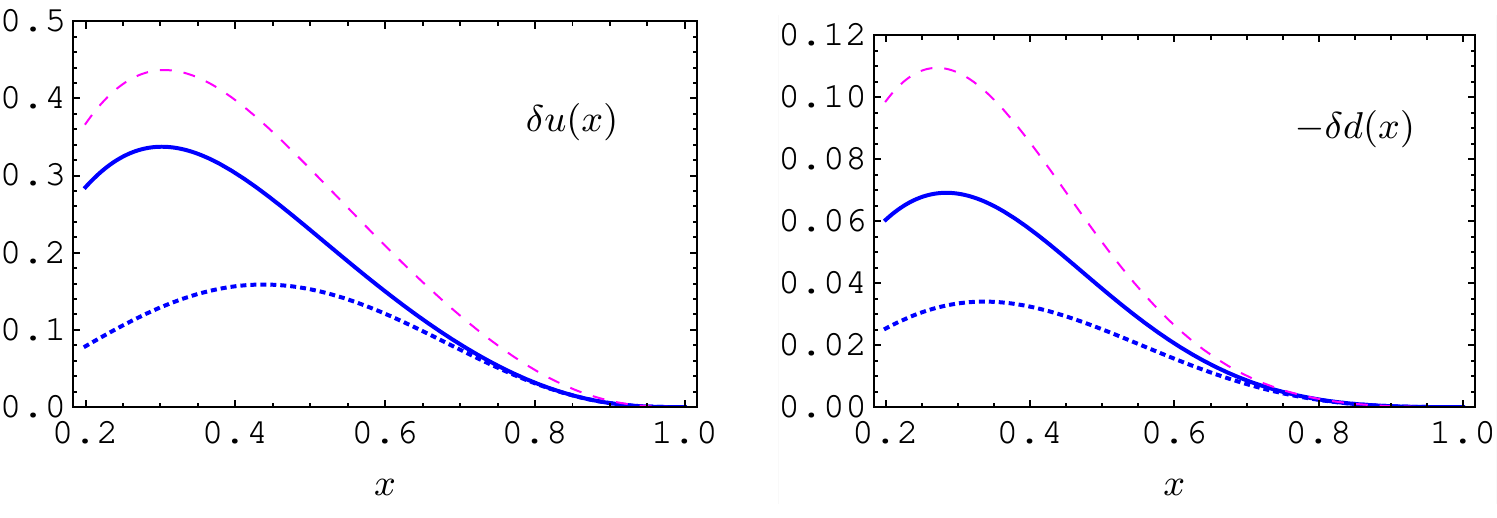}
\caption{
The quark transversity distribution $\delta q(x)$.
The solid blue curve is our model prediction taking into account the
contributions of the valence three-quark state and the state involving one additional gluon.
The contribution of the valence state alone is shown by dots for comparison.
The Soffer bound (\ref{eq:Soffer}) is indicated by the (magenta) dashed curve. }
\label{fig:transversity}
\end{center}
\end{figure*}

The results for the transversity distributions $\delta u(x)$, $\delta d(x)$
are shown in Fig.~\ref{fig:transversity}. These distributions are only very weakly
constrained by the experiment, see e.g. the discussion in
Refs.~\cite{Anselmino:2007fs,Anselmino:2007zr,Anselmino:2008sj}.
Our results are generally similar to the other existing model predictions, see
Ref.~\cite{Barone:2001sp} for a review and the corresponding references.

We remind that in this work we try to keep the model as simple as possible, restricting
ourselves to contributions of the states with total zero angular momentum and the
simplest, asymptotic shape of the four-particle quark-gluon proton distribution amplitude.
It is seen that this simple approximation captures main features of parton distributions
at large $x$ surprisingly well, although more sophisticated models are certainly needed
for a quantitative description.

%%%%%%%%%%%%%%%%%%%%%%%%%%%%%%%%%%%%%%%%%%%%%%%%%%%%%%%%%%%%%%%%%%%%%%%%%%%%%%%%%%%%%%%%%%%%%%%
\section{Twist-3 observables}
%%%%%%%%%%%%%%%%%%%%%%%%%%%%%%%%%%%%%%%%%%%%%%%%%%%%%%%%%%%%%%%%%%%%%%%%%%%%%%%%%%%%%%%%%%%%%%%

\subsection{Quark-antiquark-gluon correlation functions}

A description of twist-three observables in the framework of collinear factorization
involves
quark-antiquark-gluon correlation functions which are defined as matrix elements
of nonlocal (light-ray) three-particle operators. In the literature there exists
apparently no ``standard'' definition of such operators, and also no standard notation.
One of the usual choices~\cite{Braun:2000yi} is to consider the operators
\begin{equation}
\mathbb{S}^\pm_\mu(z_1,z_2,z_3)=
\frac12\bar q(z_1)\big[i\widetilde{F}_{\mu+}(z_2)\pm F_{\mu+}(z_2)\gamma_5\big]\gamma_+ q(z_3)\,
\label{eq:Spm}
\end{equation}
and define the twist-three correlations functions $D^\pm_q$ as the matrix elements
\begin{eqnarray}
\vev{p,s|S^\pm_\mu(z_1,z_2,z_3)|p,s}&=&4m_Ni(pn)[s_\mu(pn)-p_\mu(sn)]
\nonumber\\
&&{}\hspace*{-1.0cm}\times \int \mathcal{D}x\,  e^{ipn\sum z_i x_i} D^\pm_q(x_i)\,,
\label{eq:Dpm}
\end{eqnarray}
where $s_\mu$ is the proton spin vector which we assume to be normalized as $s^2=-1$.
This formulation is often used  e.g. in the studies of the nucleon structure function
$g_2(x,Q^2)$.

Here and below the integration measure $\mathcal{D}x$ is defined as
\begin{equation}
\int \mathcal{D}x = \int dx_1 dx_2 dx_3 \,\delta(\sum x_i)\,.
\end{equation}
The difference to $[dx]_3$ (\ref{measurek}) is that the momentum fractions sum up
to zero.

A subtlety in using this definition is that the twist-three and twist-four contributions
in $\mathbb{S}^\pm_\mu$ are not separated on the operator level. It can be more convenient
to forgo the explicit Lorentz covariance and restrict oneself to transverse spin
polarizations $(s_T\cdot n)=0$ introducing another set of operators~\cite{Braun:2009mi}:
\begin{equation}
\mathfrak{S}^\pm(z)=2is_T^\mu[\mathbb{S}^+_\mu(z_1,z_2,z_3)\pm\mathbb{S}^-_\mu(z_3,z_2,z_1)]\,.
\label{SSF}
\end{equation}
The operators $\mathfrak{S}^\pm(z)$ are even (odd) with respect to the charge conjugation.
One can show that~\cite{Braun:2009mi}
$$\left(\mathfrak{S}^{\pm}({z})\right)^\dagger= \pm\mathfrak{S}^{\pm}({z})$$
so that the $C$-even (``plus'') and $C$-odd (``minus'')
$\mathfrak{S}$-operators are hermitian and antihermitian,
respectively.

The corresponding matrix elements define the $C$-even and the $C$-odd
twist-three correlations functions
\begin{equation}
\vev{p, s_T|\mathfrak{S}^\pm({z})|p, s_T}=2(pn)^2\int \mathcal{D}x\, e^{-i(pn)\sum_k x_k z_k}
\,\mathfrak{S}^\pm({x})\,,
\label{SFM}
\end{equation}
which are related to the $D^\pm$--functions introduced above as
\begin{eqnarray}\label{DS}
8m_N D^+(x_1,x_2,x_3)&=&
\mathfrak{S}^+(x_1,x_2,x_3)-\mathfrak{S}^-(x_1,x_2,x_3)\,,
\nonumber\\
8 m_N D^-(x_1,x_2,x_3)&=&
\mathfrak{S}^+(x_3,x_2,x_1)+\mathfrak{S}^-(x_3,x_2,x_1)\,.
\nonumber\\
\end{eqnarray}
Note that we use the same notation $\mathfrak{S}^\pm$ for the operators
and the matrix elements, which hopefully will not lead to a confusion.

The helicity structure of the twist-three correlation functions can
be made explicit going over to the two-component spinor notation.
One obtains
\begin{equation}\label{Sdef}
 \mathfrak{S}^{\pm}({z}) = -{ig}\Big[\bar s\, \mathcal{Q}^{\pm}({z})
- s\, \widetilde{\mathcal{Q}}^{\pm}({z}) \Big]\,,
\end{equation}
where $s=s^{1}+is^2$, $\bar s=s^{1}-is^2$, and
\begin{eqnarray}
\mathcal{Q}^{\pm}({z})&=\!&
\bar q^{\downarrow}_+(z_1)f_{++}(z_2)q^{\downarrow}_+(z_3)
\pm \bar q^{\uparrow}_+(z_3) f_{++}(z_2)q^{\uparrow}_+(z_1)\,,
\nonumber\\
 \widetilde{\mathcal{Q}}^{\pm}({z})&=\!&
\bar q^{\uparrow}_+(z_1)\bar f_{++}(z_2)q^{\uparrow}_+(z_3)
\pm \bar q^{\downarrow}_+(z_3) \bar f_{++}(z_2)q^{\downarrow}_+(z_1)\,.
\nonumber\\
\end{eqnarray}
The nucleon state with a transverse polarization can be expressed in terms of
the helicity states $\ket{p,\pm}$ as
$$
\ket{p,s_T}=\frac{1}{\sqrt{2}}\Big[\ket{p,+}+s\,\ket{p,-}\Big].
$$
Taking into account that the operators $\mathcal{Q}^{\pm}({z})$ increase and
$\widetilde{\mathcal{Q}}^{\pm}({z})$ decrease helicity, it follows that
\begin{eqnarray}\label{}
\lefteqn{\vev{p,s_T|\mathfrak{S}^{\pm}({z})|p,s_T}=}
\nonumber\\&=&
{i g}\Big[
\vev{p,+|\mathcal{Q}^{\pm}(z)|p,-}-\vev{p,-|\widetilde{\mathcal{Q}}^{\pm}(z)|p,+}
\Big]\,.
\label{eq:SQrel}
\end{eqnarray}
It is easy to see that
$\widetilde{\mathcal{Q}}^{\pm}({z})=\pm[\mathcal{Q}^{\pm}({z})]^\dagger$, so
that the two matrix elements on the r.h.s. of Eq.~(\ref{eq:SQrel}) are related and
one does not need to consider the operators with a ``tilde'' explicitly.
%%%%%%%%%%%%%%%%%%%%%%%%%%%%%%%%%%%%%%%%%%%%%%%%%%%%%%%%%%%%%%%%%%%%%%
\begin{figure*}[t]
\begin{center}
\includegraphics[width=14.0cm,clip=true]{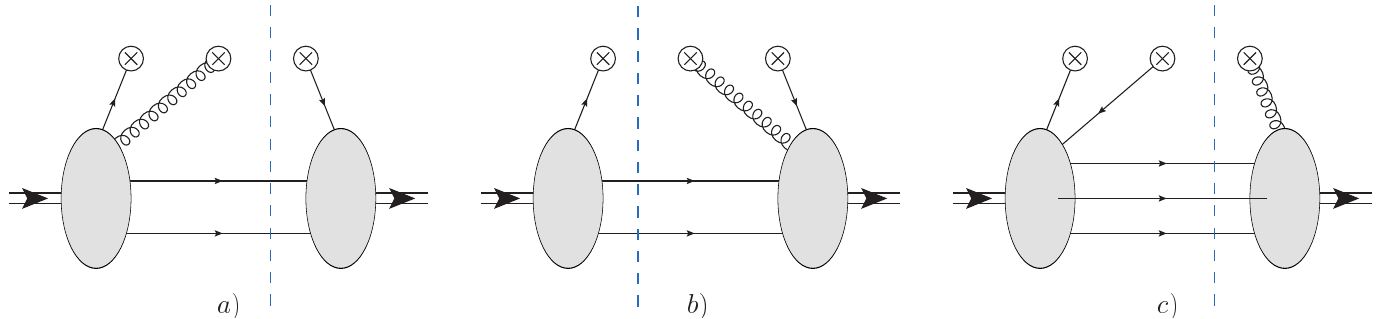}
\caption{
Twist-three correlation functions from the overlap of light-cone wave functions.
}
\label{fig:t3pd}
\end{center}
\end{figure*}
%%%%%%%%%%%%%%%%%%%%%%%%%%%%%%%%%%%%%%%%%%%%%%%%%%%%%%%%%%%%%%%

We define the distributions $\mathcal{Q}^{\pm}(x)$ as
\begin{equation}
\vev{p,+|\mathcal{Q}^{\pm}(z)|p,-}=-2i (pn)^2 \int \mathcal{D}x\,
     e^{-i(pn) \sum x_kz_k} \mathcal{Q}^{\pm}(x)\,.
\label{eq:Qpm}
\end{equation}
The $P$-parity implies (cf.~\cite{Braun:2009mi})
\begin{equation}
\bigl(\mathcal{Q}^{\pm}(-x)\bigr)^*=\pm\mathcal{Q}^{\pm}(x)
\label{eq:parity}
\end{equation}
and finally
\begin{equation}
\label{SSg}
\mathfrak{S}^{\pm}({x}) = - g\,\mathcal{Q}^{\pm}(x)\,.
\end{equation}

In the light-cone formalism, twist-three correlation functions are
generated by the interference of Fock states with different particle content as
illustrated in Fig.~\ref{fig:t3pd}.
The contributions shown schematically
in Fig.~\ref{fig:t3pd}a,b correspond to the
interference of the  three-quark and three-quark-gluon wave functions,
whereas the one in Fig.~\ref{fig:t3pd}c stands for the interference
of the three-quark-gluon state with the one containing an extra
quark-antiquark pair.
The latter term contributes to a different kinematic region in momentum
fractions compared to the first two terms and is missing to our accuracy.

Explicit expressions for the three-quark and three-quark-gluon Fock states
for the nucleon with positive helicity are given in Sec.~\ref{sect:light-cone}.
The corresponding states for the nucleon of negative helicity are given
by the same expressions, Eqs.~(\ref{Ansatz}),~(\ref{gminus}),~(\ref{gplus}),
where helicities of creation operators have to be flipped. The wave functions
of the three-quark-gluon states of the nucleon with positive and negative
helicity are the same,
$[\Psi^{\uparrow\downarrow}_{1234}]_{(-)}=[\Psi^{\uparrow\downarrow}_{1234}]_{(+)}$,
whereas for the valence three-quark state there is an overall sign difference:
$[\Psi^{(0)}_{123}]_{(-)}= - [\Psi^{(0)}_{123}]_{(+)}$.
All matrix elements in question can be expressed in terms of two
correlation functions $\mathcal{Q}^{\uparrow(\downarrow)}_q(x)$ defined as
\begin{align}\label{qpm1}
&{}_{uud}\vev{p,+|\bar q^{\uparrow}_+(z_3)f_{++}(z_2)q^{\uparrow}_+(z_1)|p,-}_{uudg^\uparrow}
\notag\\
&\hspace*{2cm}=-2ip_+^2
\int \mathcal{D}x\, e^{-ip_+\sum x_iz_i} \mathcal{Q}^{\uparrow}_q(x)\,,
\notag\\
&{}_{uud}\vev{p,+|\bar q^{\downarrow}_+(z_1)f_{++}(z_2)q^{\downarrow}_+(z_3)|p,-}_{uudg^\uparrow}
\notag\\
&\hspace*{2cm}=-2ip_+^2
\int \mathcal{D}x\, e^{-ip_+\sum x_iz_i} \mathcal{Q}^{\downarrow}_q(x)\,,
\end{align}
where the subscript $q=u,d$ stands for quark flavor.
In particular
\begin{align}
\mathcal{Q}_q^\pm(x)=&\mathcal{Q}_q^\downarrow(x)\pm \mathcal{Q}_q^\downarrow(-x)
+ \mathcal{Q}_q^\uparrow(-x) \pm \mathcal{Q}_q^\uparrow(x)\,.
\label{eq:Qupdown}
\end{align}

Using the ansatz for the LCWFs in Eqs.~(\ref{Psi0}),(\ref{PhiPsi})
one can represent $\mathcal{Q}_q^{\uparrow(\downarrow)}(x)$ as convolution integrals of the
distribution amplitudes. We obtain
\begin{widetext}
\begin{align}\label{Qupdown:exp}
\mathcal{Q}^\downarrow_d(x)=&\frac12 A\theta(-x_1,x_2,x_3)\frac{1}{x_1}
\int\frac{d\xi_1}{\xi_1}\frac{d\xi_2}{\xi_2}\delta(1+x_1-\xi_1-\xi_2)
\Big[\phi(\xi_1,-x_1,\xi_2)+\phi(\xi_2,-x_1,\xi_1)\Big]
\psi_g(\xi_1,x_3,\xi_2,x_2)\,,
\notag\\
\mathcal{Q}^\downarrow_u(x)=&\frac12 A\theta(-x_1,x_2,x_3)\frac{1}{x_1}
\int\frac{d\xi_1}{\xi_1}\frac{d\xi_2}{\xi_2}\delta(1+x_1-\xi_1-\xi_2)
\phi(\xi_1,-x_1,\xi_2)\Big[
\psi^{(1)}_g(\xi_1,x_3,\xi_2,x_2)-\psi^{(2)}_g(\xi_1,\xi_2,x_3,x_2)\Big],
\notag\\
\mathcal{Q}^\uparrow_d(x)=&A\theta(x_1,x_2,-x_3)\frac{1}{x_3}
\int\frac{d\xi_1}{\xi_1}\frac{d\xi_2}{\xi_2}\delta(1+x_3-\xi_1-\xi_2)
\phi(\xi_1,\xi_2,-x_3)\left[
\frac12\psi^{(1)}_g(\xi_1,\xi_2,x_1,x_2)+\psi^{(2)}_g(\xi_1,x_1,\xi_2,x_2)\right],
\notag\\
\mathcal{Q}^\uparrow_u(x)=&-A\theta(x_1,x_2,-x_3)\frac{1}{x_3}
\int\frac{d\xi_1}{\xi_1}\frac{d\xi_2}{\xi_2}\delta(1+x_3-\xi_1-\xi_2)
\Big\{\phi(-x_3,\xi_1,\xi_2)\left[
\psi^{(1)}_g(x_1,\xi_1,\xi_2,x_2)+\frac12\psi^{(2)}_g(x_1,\xi_2,\xi_1,x_2)\right]
\notag\\
&+\Big[\phi(-x_3, \xi_2,\xi_1)+\phi(\xi_1,\xi_2,-x_3)\Big]\Big[
\psi_g(x_1,\xi_2,\xi_1,x_2)-\frac12
\psi_g(\xi_1,\xi_2,x_1,x_2)\Big]\Big\},
\end{align}
\end{widetext}
where it is implicitly assumed that $x_1+x_2+x_3=0$,
the Heaviside step-function with several arguments is defined as
$\theta(a,b,c)\equiv\theta(a)\theta(b)\theta(c)$, and
\begin{equation}
A=\frac13 (4\pi)^4\left(\frac{a_3^2a_g^2}{a_3^2+a_g^2}\right)^2.
\label{eq:A}
\end{equation}
The distribution $\psi_g$ is defined in Eq.~(\ref{def:varphi}).
The QCD sum rule result $\lambda_3^g \ll \lambda_{1,2}^g$ (\ref{eq:lambdag}) implies that
$\psi_g^{(1)}\simeq \psi_g^{(2)}$ and as a consequence both helicity
down functions $\mathcal{Q}_{u,d}^\downarrow(x)$ are suppressed in comparison with the
helicity up functions $\mathcal{Q}_{u,d}^\uparrow(x)$.
\begin{figure}[t]
\includegraphics[width=8cm]{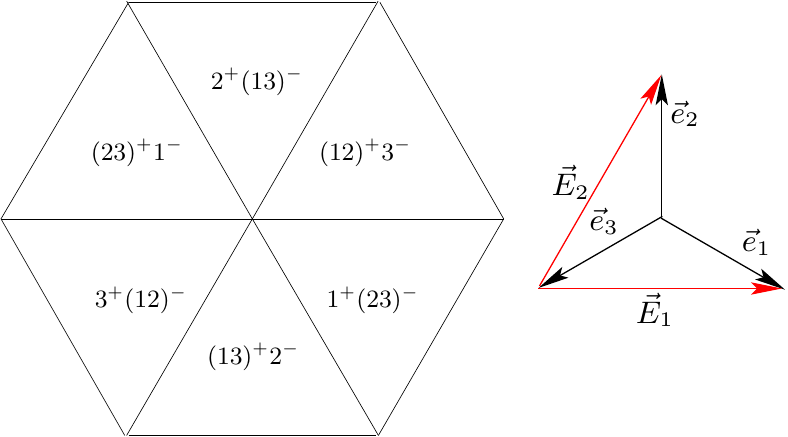}
\caption{Support properties of twist-three correlation functions
in barycentric coordinates.
For the explanation of different regions see text.
\label{fig:support}}
\end{figure}
Note that we use a symmetric notation with quark, antiquark
and gluon momentum fractions are treated equally so that the momentum
conservation condition is $x_1+x_2+x_3=0$.
Support properties of the correlation
functions~\cite{Jaffe:1983hp} can most easily be shown going over to
barycentric coordinates~\cite{Braun:2009mi} as shown in
Fig.~\ref{fig:support}:
\begin{align*}\label{bary}
\vec{x}=x_1\vec{e}_1+x_2\vec{e}_2 + x_3\vec{e}_3 = x_1\vec{E}_1+x_2\vec{E}_2\,.
\end{align*}
Three-parton correlation functions, in general, ``live''
inside a hexagon-shaped area which can further be decomposed in six
different regions (triangles). The triangles labeled
$(12)^+3^-$, $2^+(13)^-$, etc., correspond to different subprocesses
at the parton level~\cite{Jaffe:1983hp}; For each parton $k=1,2,3$\,
``plus'' stays for emission ($x_k>0$) and ``minus'' for absorption ($x_k<0$).
Alternatively, one may think of
``plus'' and ``minus'' labels as indicating whether the corresponding parton
appears in the direct or the final amplitude in the cut diagram,
cf.~Fig.~\ref{fig:t3pd}.
It is important that different regions do not have autonomous scale dependence;
they ``talk'' to each other and get mixed under the evolution,
see Ref.~\cite{Braun:2009mi} for a detailed discussion.

\begin{figure*}[t]
\begin{center}
\includegraphics[width=17cm,clip=true]{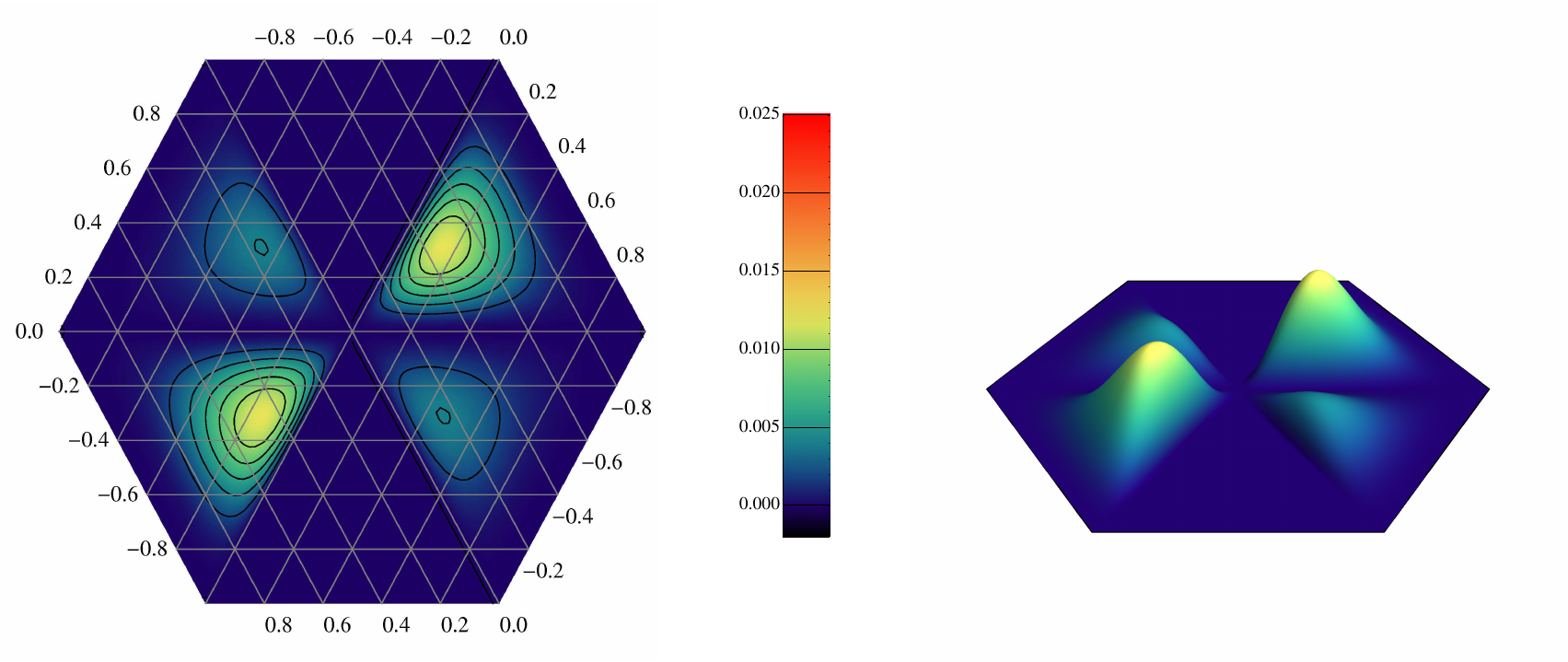}
\end{center}
\caption{
The quark-antiquark-gluon twist-three correlation function $\mathcal{Q}_{d}^+(x)$
at the reference scale 1 GeV.
}
\label{fig:Qdplus}

\begin{center}
\includegraphics[width=17cm,clip=true]{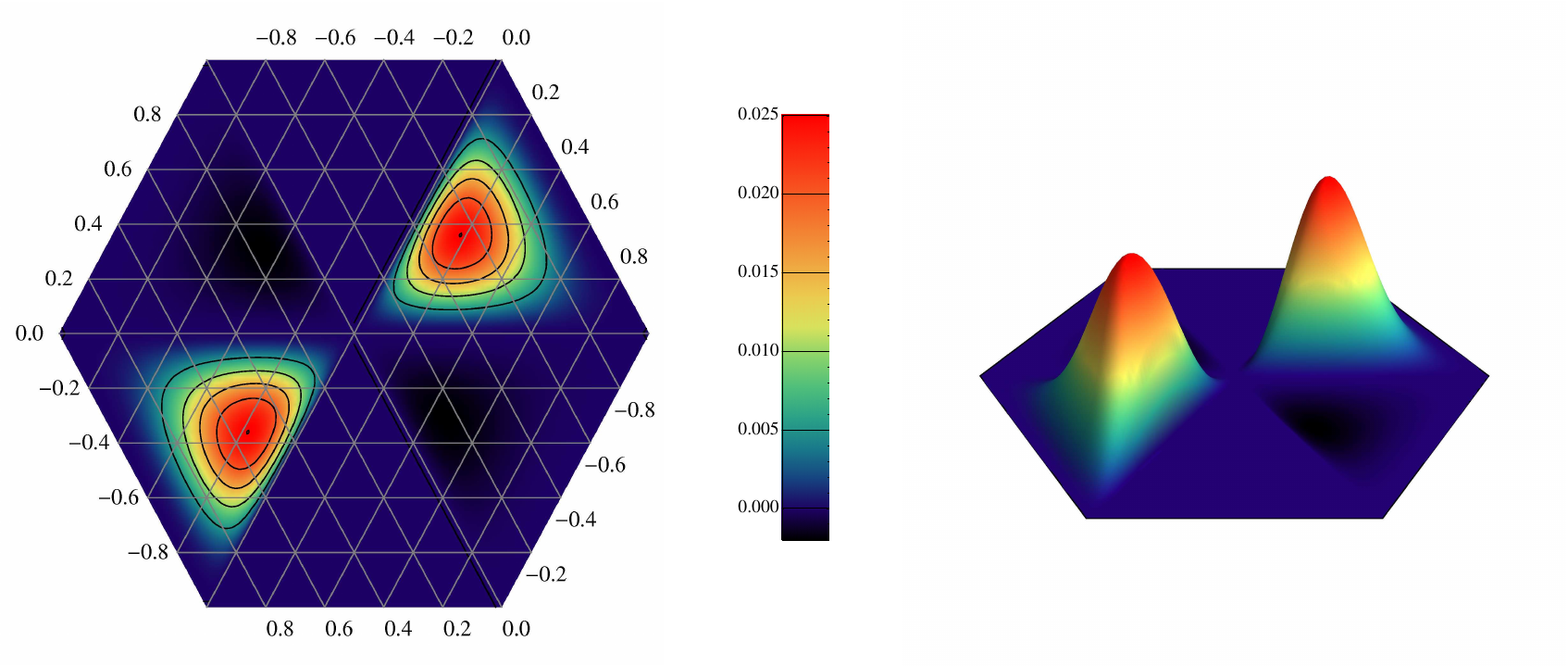}\hspace*{25mm}
\end{center}
\caption{
The quark-antiquark-gluon twist-three correlation function $-\mathcal{Q}_{u}^{+}(x)$
at the reference scale 1 GeV.
}
\label{fig:Quplus}
\end{figure*}

Our model predictions for the correlation functions $\mathcal{Q}_{d}^+(x)$
and  $-\mathcal{Q}_{u}^{+}(x)$ (note opposite sign), Eq.~(\ref{eq:Qpm}),
are shown in Fig.~\ref{fig:Qdplus} and Fig.~\ref{fig:Quplus}, respectively.
Both distributions are symmetric with respect to the center
of the hexagon: $\mathcal{Q}_q^+(x_1,x_2,x_3) = \mathcal{Q}_q^+(-x_1,-x_2,-x_3)$,
which is a consequence of $P$-parity, cf. Eq.~(\ref{eq:parity}).
Each of the four terms  $\mathcal{Q}_{q}^{\uparrow(\downarrow)}(\pm x)$ in Eq.~(\ref{eq:Qupdown})
is confined to a different ``triangle'' and, hence, has a different partonic interpretation:
\begin{eqnarray}
\mathcal{Q}_{q}^{\uparrow}(x)  \,:\, (12)^+3^-,
&~~&
\mathcal{Q}_{q}^{\uparrow}(-x) \,:\, 3^+(12)^-,
\nonumber\\
\mathcal{Q}_{q}^{\downarrow}(x) \,:\, (23)^+1^-,
&~~&
\mathcal{Q}_{q}^{\downarrow}(-x) \,:\, 1^+(23)^-.
\end{eqnarray}
The larger contributions, e.g. in the $(12)^+3^-$ region, correspond to (valence) quark
emission with momentum fraction $x_1>0$ and subsequent absorption with momentum fraction $-x_3 > x_1>0$,
accompanied with gluon emission with momentum fraction $x_2>0$.
The smaller contributions, e.g. in the $1^+(23)^-$ region, differ from the above
in that the gluon with momentum fraction $-x_2>0$ is absorbed and thus $x_1 > -x_3 > 0$.
Note that there is no symmetry between
gluon emission and absorption, which may be somewhat
counterintuitive.

The dominant, gluon emission contribution to the $\bar u G u $ correlation function $\mathcal{Q}_{u}^{\uparrow}(x)$
is roughly factor two larger compared to the $\bar d G d $ distribution, $\mathcal{Q}_{d}^{\uparrow}(x)$,
and has the opposite sign. The contributions of gluon absorption,
$\mathcal{Q}_{u}^{\downarrow}(x)$ and $\mathcal{Q}_{d}^{\downarrow}(x)$, have the same sign for $u$- and $d$-quarks, and
are much smaller compared to gluon emission.

Our model correlation functions vanish in the $2^+(13)^-$ and $(13)^+2^-$
regions. This property is an artefact of neglecting contributions of the
type shown in Fig.~\ref{fig:t3pd}c which are formally higher order in the
Fock expansion. These contributions can be estimated using a model
for the five-parton
$qqq(\bar q q)$ state from Ref.~\cite{Diehl:1998kh} and turn out
to be considerably smaller than the ones considered here.

The ``minus'' correlation functions
$\mathcal{Q}_{u}^{-}(x)$ and $\mathcal{Q}_{d}^{-}(x)$
are obtained from the ``plus'' ones by changing the sign of the contributions in the
$(12)^+3^-$ and $1^+(23)^-$ regions, so we do not show them separately.

\subsection{The structure function $g_2(x,Q^2)$}

The structure function $g_2(x_B, Q^2)$ is given by the sum of the Wandzura-Wilczek (WW)
and genuine twist-3 contributions
\begin{equation}
g_2(x_B,Q^2) = g^{WW}_2(x_B,Q^2)+g^{tw-3}_2(x_B,Q^2)\,.
\end{equation}
The WW contribution reads
\begin{equation}
g^{WW}_2(x_B,Q^2)=-g_1(x_B,Q^2)+\int_{x_B}^1\frac{dy}{y}g_1(y,Q^2)\,,
\label{eq:WW}
\end{equation}
where
\begin{equation}
g_1(x_B,Q^2)=\frac12\sum_{q} e_q^2\left[\Delta q(x_B,Q^2)+\Delta q(-x_B,Q^2)\right]\,.
\end{equation}
The twist-3 contribution $g^{tw-3}_2(x_B,Q^2)$ can be written as
\begin{eqnarray}
\lefteqn{g^{tw-3}_2(x_B, Q^2)=}
\nonumber\\
&=&\frac12 \sum_q e_q^2
\int_{x_B}^1 \frac{d\xi}{\xi}\Big[\Delta q_T(\xi,Q^2)+\Delta q_T(-\xi,Q^2)\Big]\,,
\end{eqnarray}
where $\Delta q_T(\xi)$ is defined in terms of the $D^+$--function
introduced in Eq.~(\ref{eq:Dpm}):
\begin{equation}
\Delta q_T(\xi)=4\int \mathcal{D}x\, D_q^+(x)\frac{d}{dx_3}\left[
\frac{\delta(\xi+x_3)-\delta(\xi-x_1)}{x_1+x_3}\right]\,.
\label{eq:qT}
\end{equation}
As above, the subscript $q$ refers to the contribution of a given quark flavor.
In terms of the $\mathcal{Q}_q^{\uparrow(\downarrow)}(x)$--functions one obtains
\begin{align}\label{DQ}
D_q^+(x)=&-\frac{g}{4m_N}\left[\mathcal{Q}_q^\uparrow(x)+\mathcal{Q}_q^\downarrow(-x)\right].
\end{align}
To avoid confusion, in this section we use the notation $x_B$ for the Bjorken variable,
whereas $x$ is reserved for the set of parton momentum fractions $x=\{x_1,x_2,x_3\}$.

\begin{figure*}[t]
\begin{center}
\includegraphics[width=17cm,clip=true]{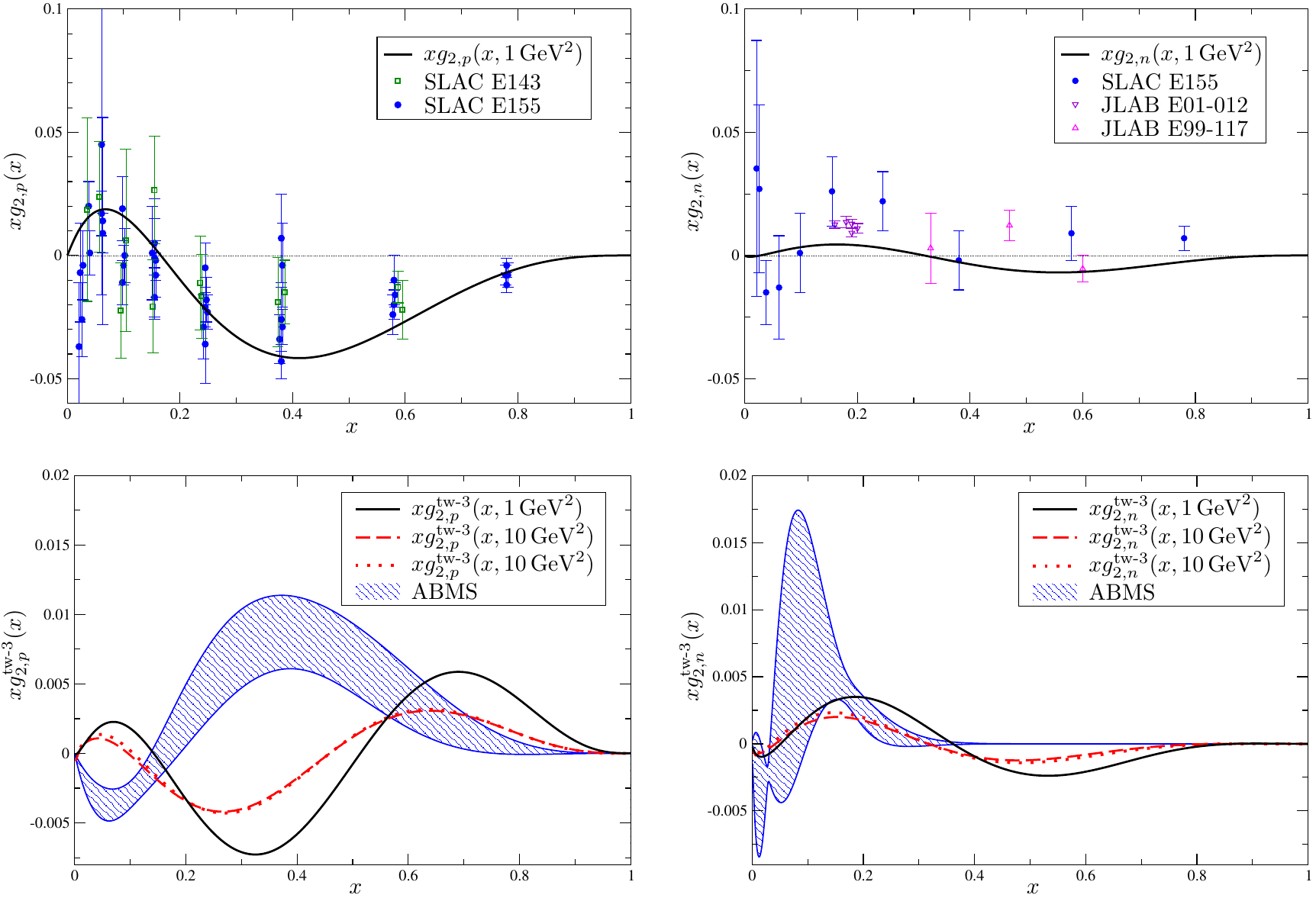}
\end{center}
\caption{\small\sf
Upper panels: Experimental results on the proton (left) and neutron (right)
structure function $g_2(x_B,Q^2)$ compared to our model calculation
at the scale $Q^2=1$~GeV$^2$.
Lower panels: The twist-three contributions  $x g^{twist-3}_2(x_B,Q^2)$
for the proton (left) and neutron (right) compared to the analysis
in Ref.~\cite{Accardi:2009au} (shaded areas). Our model predictions
at the scale $Q^2=1$~GeV$^2$ and $Q^2=10$~GeV$^2$ are shown by the black solid and
dashed red curves, respectively. The predictions at $10$~GeV$^2$ obtained using
an approximate evolution equation from Refs.~\cite{Ali:1991em,Braun:2001qx}
are shown by the red dotted curves for comparison.
}
\label{fig:g2}
\end{figure*}

Under a plausible assumption that the spin-dependent part of the
forward Compton amplitude satisfies a dispersion relation without subtractions,
the integral of $g_2(x_B, Q^2)$ and, hence, of $g^{tw-3}_2(x_B, Q^2)$
vanishes~\cite{Burkhardt:1970ti}
\begin{equation}
 \int_0^1 dx_B\, g^{tw-3}_2(x_B, Q^2) = 0\,.
\label{eq:BK}
\end{equation}
This statement is known as the Burkhardt-Cottingham (BC) sum rule.

Using Eqs.~(\ref{Qupdown:exp}) one finds that in our model $D_q^+(x)$ is nonzero only when
$x_1\geq 0$ and $x_3\leq 0$.
This, in turn, implies that $\Delta q_T(\xi)$ vanishes for $\xi<0$
(i.e. there is no antiquark contribution).
As a consequence, in our model $g^{tw-3}_2(x_B, Q^2)$ satisfies in addition to
Burkhardt-Cottingham (\ref{eq:BK}) also the Efremov-Leader-Teryaev (ELT)
sum rule~\cite{Efremov:1996hd}:
\begin{equation}
 \int_0^1 dx_B\, x_B\, g^{tw-3}_2(x_B, Q^2) = 0\,.
\label{eq:ELT}
\end{equation}

{}For the second moments one obtains
\begin{eqnarray}
 d_{2,p}&=&3\int_0^1 dx_B\,x_B^2 g_{2,p}^{tw-3}(x_B)
\nonumber\\
&=&\frac{5}{32} A f_N \Big[\lambda_2^g\left(1+\frac{5a+b}{12}\right)
+\lambda_3^g\Big(1+\frac{a+5b}{12}\Big)\Big],
\nonumber\\
 d_{2,n}&=&3\int_0^1 dx_B\,x_B^2g_{2,n}^{tw-3}(x_B)
\nonumber\\
&=&-\frac{5}{32}Af_N\Big[\lambda_2^g\Big(1+\frac{b-5a}{12}\Big)
       +\lambda_3^g\Big(1+\frac{a-5b}{12}\Big)\Big].
\nonumber\\
\end{eqnarray}
The corresponding numerical values are, at the scale 1~GeV:
\begin{eqnarray}
 d_{2,p} &=&\phantom{-}0.0016\,, \qquad
 d_{2,n} \,=\, -0.00072\,.
\end{eqnarray}
Both numbers compare very well to the lattice QCD \cite{Gockeler:2000ja},
 QCD sum rules \cite{Balitsky:1989jb,Stein:1994zk}
and chiral quark soliton model \cite{Wakamatsu:2000ex} calculations.
The negative value of $d_{2}$ for the neutron (in all models) is in conflict,
however, with the existing experimental average:
\begin{eqnarray}
 d^{\rm exp}_{2,p} &=& 0.0032 \pm 0.0017 ~\text{\cite{Anthony:2002hy}}\,,
\nonumber\\
 d^{\rm exp}_{2,n} &=& 0.0062 \pm 0.0028 ~\text{\cite{Zheng:2004ce}}\,.
\end{eqnarray}

Further, a straightforward calculation gives
\begin{eqnarray}
 g^{tw-3}_{2,p}(x_B) &=& 0.0436772\Big(\ln x_B+\bar x_B + 1/2\bar x_B^2\Big)
\nonumber\\&&\hspace*{-2.3cm}
{}+ \bar x_B^3\Big(1.57357-5.94918\bar x_B +6.74412\bar x_B^2-2.19114\bar x_B^3\Big),
\nonumber\\
 g^{tw-3}_{2,n}(x_B) &=& 0.0655158\Big(\ln x_B+\bar x_B + 1/2\bar x_B^2\Big)
\nonumber\\&&\hspace*{-2.3cm}
{}+ \bar x_B^3\Big(0.130996-1.12101\bar x_B +2.31342\bar x_B^2-1.20598\bar x_B^3\Big)
\nonumber\\
\end{eqnarray}
(at the reference scale 1 GeV) for the proton and neutron, respectively.

Our results for the full structure function $g_{2}(x_B,Q^2)$ are
compared to the experimental data \cite{Abe:1998wq,Zheng:2004ce,Kramer:2005qe}
in Fig.~\ref{fig:g2} (upper panels) and, separately,
for the twist-three contribution $g_{2}^{tw-3}(x_B)$ to the analysis in
Ref.~\cite{Accardi:2009au} (lower panels).
The twist-three contributions are shown at the model scale $Q^2=1$~GeV$^2$ and after the
evolution to a higher scale $Q^2=10$~GeV$^2$. The scale dependence was calculated
in two ways: using exact (one-loop) evolution equations for the relevant
quark-antiquark-gluon correlation functions from Ref.~\cite{Braun:2009mi} (dashed curves), and
using the much simpler evolution equation from Refs.~\cite{Ali:1991em,Braun:2001qx} which is based
on the large-$N_c$ and large-$x_B$ approximation and only
involves the $g_{2}^{tw-3}(x_B)$ structure function itself (dotted curves).
Since we are interested primarily in the large $x_B$ region, we used flavor-nonsinglet
evolution equations which are simpler.
The results
of both approaches almost coincide within the line thickness. A good accuracy of
this approximation was expected but has never been checked in a dynamical model calculation.
Note that effects of the evolution are generally significant because of large anomalous
dimensions of twist-three operators,
and have to be taken into account in the analysis of the experimental data.

As seen from Fig.~\ref{fig:g2}, the twist-three contribution to the structure function
$g_2(x_B,Q^2)$ at large $x_B$ proves to be
positive for the proton and negative for the neutron. This prediction can be traced to the
relative signs of the three-quark-gluon couplings and is largely model-independent.
It is in agreement with Ref.~\cite{Accardi:2009au}.
In the intermediate region $0.2 < x_B < 0.5$ the twist-three contribution
$g_{2,p}^{tw-3}(x_B)$ changes sign and becomes negative in our calculation,
whereas it remains positive according to the data analysis in Ref.~\cite{Accardi:2009au}.
This difference may well be due to contributions of higher Fock states, with two or more
gluons, and probably also partially remedied by using a more sophisticated model for the
three-quark-gluon wave function. A detailed analysis would be interesting but
goes beyond the tasks of this paper.
Another issue is that in~\cite{Accardi:2009au} the ELT sum rule is strongly violated,
which suggests the existence of a large positive flavor-singlet contribution at $x_B \sim 0.1$
due to gluons or sea quark-antiquark pairs. Such contributions are related to the twist-three
three-gluon correlation functions and are missing in our present framework.

\subsection{Single spin asymmetries}

The quark-antiquark-gluon correlation functions considered in this work are precisely
those responsible for transverse single spin asymmetries (SSA) observed in different
hadronic reactions, if described in the framework of collinear factorization~%
\cite{Efremov:1981sh,Efremov:1984ip,Qiu:1991pp,Qiu:1991wg,Efremov:1994dg,Qiu:1998ia,%
      Kanazawa:2000hz,Kanazawa:2000cx,Eguchi:2006mc,Koike:2007rq,Kang:2008ih}.
The distributions $T_{\bar q F q}({x})$, $\Delta T_{\bar q F q}({x})$ introduced in this context in
Ref.~\cite{Braun:2009mi} are expressed in terms of $\mathcal{Q}_q^{\uparrow(\downarrow)}$--functions as follows:
\begin{eqnarray}
\lefteqn{\hspace*{-0.5cm}T_{\bar q F q}({x_1,x_2,x_3})=}
\nonumber\\
&=&\frac14\Big[(1\!+\!\mathcal{P}_{13})\mathfrak{S}^+({x})+
(1\!-\!\mathcal{P}_{13})\mathfrak{S}^-({x})\Big]
\nonumber\\
&=&-\frac{g}2\Big[\mathcal{Q}_q^{\uparrow}(x_3,x_2,x_1)+\mathcal{Q}_q^{\uparrow}(-x_1,-x_2,-x_3)
\nonumber\\
&&{}
+\mathcal{Q}_q^{\downarrow}(x_1,x_2,x_3)+\mathcal{Q}_q^{\downarrow}(-x_3,-x_2,-x_1)\Big]\,,
\nonumber\\
\lefteqn{\hspace*{-0.5cm}\Delta T_{\bar q F q}({x_1,x_2,x_3})=}
\nonumber\\
&=&-\frac14\Big[(1\!-\!\mathcal{P}_{13})\mathfrak{S}^+({x})+
(1\!+\!\mathcal{P}_{13})\mathfrak{S}^-({x})\Big]
\nonumber\\
&=&-\frac{g}2\Big[\mathcal{Q}_q^{\uparrow}(x_3,x_2,x_1)-\mathcal{Q}_q^{\uparrow}(-x_1,-x_2,-x_3)
\nonumber\\
&&{}-\mathcal{Q}_q^{\downarrow}(x_1,x_2,x_3)+\mathcal{Q}_q^{\downarrow}(-x_3,-x_2,-x_1)\Big]\,.
\end{eqnarray}
A common notation~\cite{Kang:2008ey} is to show quark momenta only:
\begin{eqnarray}
{\mathcal{T}}_{q,F}(x,x')&\equiv&
T_{\bar q F q}(-x',x'-x,x)\,,
\notag\\
{\mathcal{T}}_{\Delta q,F}(x,x')&\equiv&  \Delta T_{\bar q F q}(-x',x'-x,x)\,,
\end{eqnarray}
Written in this way, the distributions are symmetric (antisymmetric) functions of the arguments:
${\mathcal{T}}_{q,F}(x,x')={\mathcal{T}}_{q,F}(x',x)$ and
$\Delta{\mathcal{T}}_{q,F}(x,x')=-\Delta{\mathcal{T}}_{q,F}(x',x)$.
A yet another notation for the same functions in a different normalization is used in the
recent analysis in Ref.~\cite{Kanazawa:2010au}:
\begin{eqnarray}
   G_F^q(x,x') &\equiv& -\frac{2}{m_N}T_{\bar q F q}(-x',x'-x,x)\,,
\nonumber\\
  \widetilde G_F^q(x,x') &\equiv& \frac{2}{m_N} \Delta T_{\bar q F q}(-x',x'-x,x)\,.
\end{eqnarray}

\begin{figure*}[t]
\begin{center}
\includegraphics[width=17cm,clip=true]{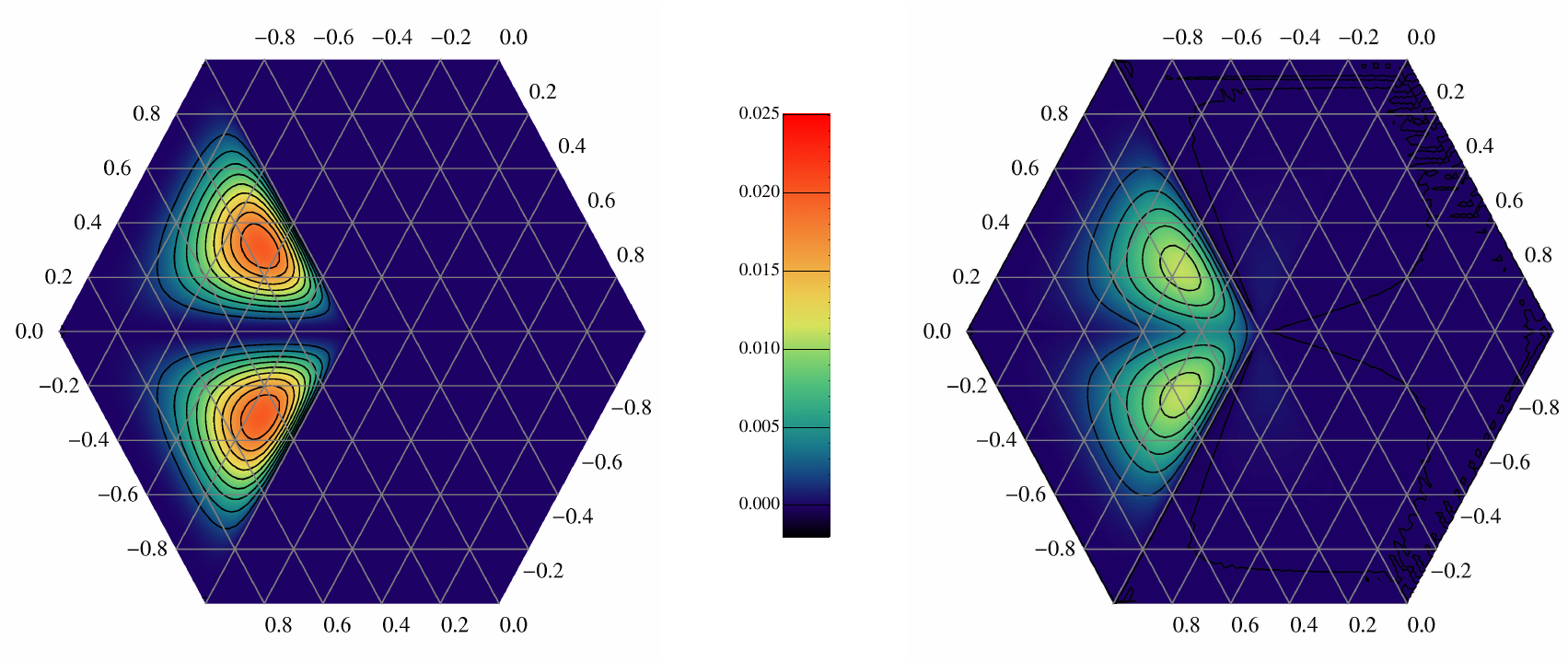}
\end{center}

\caption{
The quark-antiquark-gluon twist-three correlation function $-T_{\bar d F d}(x)$
at the reference scale $\mu^2=1$~GeV$^2$ (left) and $\mu^2=10$~GeV$^2$ (right).
}
\label{fig:TdFd}
\end{figure*}

In the framework of collinear factorization, SSAs originate from imaginary (pole)
parts of propagators in the hard coefficient functions. In the leading order,
taking a pole part enforces vanishing of one of the momentum fractions in the twist-3
parton distribution, and are classified as soft gluon pole (SGP) or soft
fermion pole (SFP), depending on which momentum is put to zero, respectively.
Such ``pole'' contributions are therefore considered to be main source of the
observed asymmetries and can be estimated from the available experimental
data~\cite{Kouvaris:2006zy,Kanazawa:2010au}.

Since our approximation for the nucleon wave function does not contain antiquarks,
the $T_{\bar q F q}$, $\Delta T_{\bar q F q}$ distributions are nonzero
in the $(23)^+ 1^-$ and $(12)^- 3^+$ regions only, cf. Fig.~\ref{fig:support}.
Moreover, both distributions vanish at the boundaries of parton regions
where one of the momentum fractions goes to zero, and, hence,
both SGP and SFP terms vanish as well. This property is an obvious
artefact of the truncation of the Fock expansion to a few lowest components:
The LCWF of each Fock state vanishes whenever momentum fraction of any parton goes to zero
and the same property holds true for the correlation functions.
Our model for the gluon distribution $x g(x)$ in Fig.~\ref{fig:PD} vanishes at $x\to 0$
for the very same reason.

For the leading-twist parton distributions, a possible way out is to assume the valence-type
input at a certain low scale, and construct realistic dynamical models by applying
QCD evolution equations that include multiple soft gluon radiation. This approach was
suggested by Gl{\"u}ck, Reya and Vogt (GRV) \cite{Gluck:1994uf,Gluck:1998xa} and proved to
be  very successful phenomenologically.
Exploiting the same idea for the twist-three distributions suggests itself.

\begin{figure*}[t]
\begin{center}
 \includegraphics[width=17cm,clip=true]{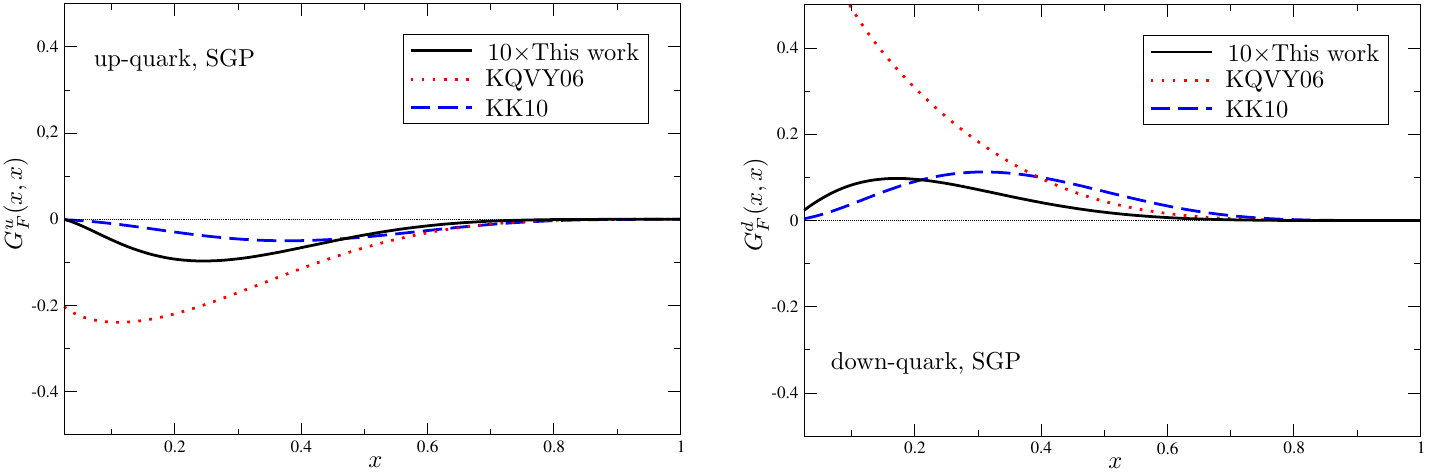}
\end{center}
\caption{\small\sf
  Radiatively generated SGP distributions $G_F^q(x,x)$ at $Q^2=10$~GeV$^2$
  rescaled by factor 10, shown by the solid curves, as
  compared to the phenomenological studies of spin asymmetries in high transverse momentum
  meson production in $pp$ colisions~\cite{Kouvaris:2006zy,Kanazawa:2010au}.
}
\label{fig:SP}
\end{figure*}

It is easy to see that both the SGP and SFP contributions reappear once QCD
evolution is taken into account. The full one-loop evolution equation for the
functions $T_{\bar q F q}$, $\Delta T_{\bar q Fq}$ is rather cumbersome and can be found
in~\cite{Braun:2009mi}. For our present purposes the flavor-nonsinglet evolution equation
is sufficient. Restricting ourselves to the SGP kinematics $x_2\to 0$ one obtains,
to the one-loop accuracy
\begin{widetext}
\begin{eqnarray}\label{KQe}
\mathcal{T}_{q,F}(x,x;\mu^2)&=& \mathcal{T}_{q,F}(x,x;\mu_0^2)+
\frac{\alpha_s}{2\pi}\ln\frac{\mu^2}{\mu_0^2}\Big\{\int_x^1 \frac{d\xi}{\xi}\Big[P_{qq}(z) \mathcal{T}_{q,F}(\xi,\xi)
+\frac{N_c}{2}\frac{1+z}{1-z}\mathcal{T}_{q,F}(x,\xi)
-\frac{N_c}{2}\frac{1+z^2}{1-z}\mathcal{T}_{q,F}(\xi,\xi)
\nonumber\\&&{}
-\frac{N_c}{2}\mathcal{T}_{\Delta q,F}(x,\xi)
+\frac1{2N_c} (1-2z)\mathcal{T}_{q,F}(x,x-\xi)
- \frac1{2N_c}\mathcal{T}_{\Delta q,F}(x,x-\xi)
\Big] -N_c \mathcal{T}_{q,F}(x,x) \Big\}^{\mu_0^2},
\end{eqnarray}
\end{widetext}
where it is assumed that $x>0$, $P_{qq}(z)$ is the usual Altarelli--Parisi splitting function
and $z=x/\xi$. Even if $\mathcal{T}_{q,F}(x,x;\mu_0^2)=0$, a non-zero SGP contribution is
generated at a higher scale $\mu^2$. It is given by a certain integral of
$T_{\bar q F q}$, $\Delta T_{\bar q Fq}$ away from the line $x_2=0$, and involves large quark momentum
fractions only, $\xi > x$. (For a detailed discussion of integration regions in Eq.~(\ref{KQe}) see
Ref.~\cite{Braun:2009mi}.)

One difficulty in following the GRV approach is that the initial condition for the evolution
has to be taken at a very low scale $\mu^2_{\rm GRV}\simeq 0.25$~GeV$^2$~\cite{Gluck:1994uf,Gluck:1998xa}
whereas our model is formulated at $\mu^2_0 = 1$~GeV$^2$. The advantage of using the higher scale
is that we have been able to use QCD perturbation theory and operator product expansion
to get some insight in the structure of the lowest Fock states, but the price to pay is
that the nucleon at the scale 1~GeV already contains significant admixture of yet higher states,
with several gluons and quark-antiquark pairs, which we do not know much about.
These additional contributions are not taken
into account in this work, and this is the reason that we underestimate parton distributions
at small $x$, cf.~Fig.~\ref{fig:PD}.

A consistent implementation of the GRV program would require to give up QCD motivated
models for the $qqq$ and $qqqg$ states and resort to purely phenomenological
parametrizations. We leave this study for future work.
Instead, in what follows we show the results corresponding to the evolution of
our model twist-three parton distribution from $1$~GeV$^2$ to an
{\it ad hoc}\ scale $\mu^2 = 10$~GeV$^2$.
This calculation should be considered as an illustration, since effects of the QCD evolution
from the GRV scale $\mu^2_{\rm GRV}\simeq 0.25$~GeV$^2$ are, generally,  much larger.

As an example, in Fig.~\ref{fig:TdFd} we show the
the quark-antiquark-gluon twist-three correlation function $T_{\bar d F d}(x)$ (with opposite sign)
at the model scale $\mu^2=1$~GeV$^2$ (left) and after the evolution
to $\mu^2=10$~GeV$^2$ (right). As already mentioned above, in our model (left picture)
this correlation function is only nonzero in the two left-most triangle regions corresponding
to emission and subsequent absorption of the (valence) quark. The upper and the lower triangles
corresponds to gluon emission and absorption, respectively. The
longest diagonals of the hexagon, connecting diametrically opposite vertices,
correspond to vanishing of one of the parton momentum fractions.
In particular, on the horizontal diagonal $x_2=0$, i.e. it corresponds to the SGP kinematics,
and on the other two diagonals either $x_1=0$ or $x_3=0$, so they stand for the SFPs.
The two triangles that come next to the right and include the upper (or the lower)
edges of the hexagon, correspond to the contributions of the type shown in
Fig.~\ref{fig:t3pd} where a gluon is emitted
and a quark-antiquark pair is absorbed (or vice versa). These contributions are thus analogous to the
so-called ERBL regions in off-forward parton distributions and, formally, are of higher order
in the Fock expansion. Finally, the two right-most triangles correspond to the antiquark
distributions.

Once the QCD evolution is taken into account, different parton regions get mixed.
In particular the gap between the $x_2>0$ and $x_2 < 0$ regions gets closed and
the SGP term appears, see Fig.~\ref{fig:TdFd} (right picture). The SFP terms are
also generated, but remain very small because the corresponding terms in the evolution
equations are $1/N_C$ suppressed.

The radiatively generated SGP distributions $G_F^q(x,x)$
%(upper panels) and SFP distributions $G_F^q(0,x) + \widetilde G_F^u(0,x)$ (lower panels)
at $Q^2=10$~GeV$^2$ are shown in Fig.~\ref{fig:SP} and compared there with the results of
phenomenological studies of spin asymmetries in high transverse momentum
meson production in $pp$ colisions~\cite{Kouvaris:2006zy,Kanazawa:2010au}.
Our distributions are of the same sign and similar shape
compared to these studies, but about one order of magnitude smaller.
It is plausible that much larger SGP contributions can be generated from the
similar valence-like ansatz if the QCD evolution is started at a low scale
of the order of  $\mu^2_{\rm GRV}\simeq 0.25$~GeV$^2$~\cite{Gluck:1994uf,Gluck:1998xa}.
The SFP contributions that we obtain in this exercise appear to be two orders of magnitude
below the estimates in Ref.~\cite{Kanazawa:2010au}, albeit with the correct sign.
It is unlikely that such large contributions can be obtained radiatively starting
from the valence-like ansatz, unless one assumes the existence of antiquarks with large momentum
fraction at low scales in the proton WF.

%%%%%%%%%%%%%%%%%%%%%%%%%%%%%%%%%%%%%%%%%%%%%%%%%%%%%%%%%%%%%%%%%%%%%%%%%%%%%%%%%%%%%%%%%%%%%%%
\section{Conclusions}
%%%%%%%%%%%%%%%%%%%%%%%%%%%%%%%%%%%%%%%%%%%%%%%%%%%%%%%%%%%%%%%%%%%%%%%%%%%%%%%%%%%%%%%%%%%%%%%

In this work we explored the possibility to construct higher-twist parton distributions
in a nucleon at some low reference scale from convolution integrals of
the light-cone wave functions.

To this end we have studied the general structure and
introduced simple models for the four-particle
nucleon LCWFs involving three valence quarks and a gluon with
total orbital momentum zero, and estimated their normalization
(WF at the origin) using QCD sum rules. We have shown that
truncating the Fock expansion at this order, that is taking into
account valence three-quark configuration and those with one additional
gluon, provides one with a reasonable description of both
polarized and unpolarized parton densities at large values of
Bjorken variable $x\ge 0.5$.

Using this set of LCWFs,
twist-three quark-antiquark-gluon parton distributions have been
constructed as convolution integrals of $qqqg$ and valence
three-quark components, which enter the description of
many hard reactions in QCD in the framework of collinear factorization.
In particular the twist-three contribution to the polarized structure function
$g_2(x,Q^2)$ is given by a certain integral of the three-particle distribution
over the parton momentum fractions, and thus is a measure of its ``global''
properties. Our calculation correctly reproduces the sign and the order of magnitude
of the twist-3 term at large $x$, without free parameters.

Transverse single spin asymmetries, on the other hand, are sensitive to ``local''
properties of the three-particle correlation functions in specific configurations
where one of the momentum fractions vanishes.
Since our approximation for the nucleon wave function only includes a few lowest
Fock components, and since the LCWF of each Fock state vanishes whenever momentum
fraction of any parton goes to zero, both ``soft gluon pole'' and
``soft fermion pole'' terms vanish at the scale where the model is formulated.
They are, however, generated by QCD evolution that brings in multiple
soft gluon emission. Our results suggest that realistic dynamical models
of the the twist-three distributions (and the pole terms) can be obtained
following the GRV-like approach on the level of WFs, i.e. 
assuming that the nucleon state at a very low scale
can be described in terms of a few Fock components, including the valence quarks,
one additional gluon and, probably, a quark-antiquark pair, and applying QCD evolution
equations.

An obvious problem with this strategy is that the starting scale has to be chosen very
low, of the order of  $\mu^2_{\rm GRV}\simeq 0.25$~GeV$^2$~\cite{Gluck:1994uf,Gluck:1998xa},
and thus the modelling of the wave functions necessarily becomes purely
phenomenological. In spite of this, and the usual criticism of the application
of perturbative QCD evolution equations at very low scales, we believe that such an approach
has good chances to provide us with some intuition on the structure
of higher-twist parton distributions in general, which is currently not available.
This work is in progress and the results will be published elsewhere.

%%%%%%%%%%%%%%%%%%%%%%%%%%%%%%%%%%%%%%%%%%%%%%%%%%%%%%%%%%%%%%%%%%%%%%%%%%%%%%%%%%%%
\section*{Acknowledgements}

We would like to thank A.~Belitsky for discussions that initiated this study, 
and A.~Accardi and A.~Bacchetta for providing us with the
analytic expression for the twist-3 contributions obtained in Ref.~\cite{Accardi:2009au}.
V.M.~Braun gratefully acknowledges financial support by the
Yukawa Institute for Theoretical Physics, Kyoto University,
during the YIPQS international workshop ``High Energy Strong Interactions 2010''
where a part of this work was done.
The work by A.N.~Manashov was supported by the DFG grants 9209282, 9209506
and the RFFI grant 09-01-93108.

%%%%%%%%%%%%%%%%%%%%%%%%%%%%%%%%%%%%%%%%%%%%%%%%%%%%%%%%%%%%%%%%%%%%%%%%%%%%%
%%%%%%%%%%%%%%%%%%%%%%%%%%%%%   Appendix   %%%%%%%%%%%%%%%%%%%%%%%%%%%%%%%%%%
%%%%%%%%%%%%%%%%%%%%%%%%%%%%%%%%%%%%%%%%%%%%%%%%%%%%%%%%%%%%%%%%%%%%%%%%%%%%%

\appendix
\renewcommand{\theequation}{\Alph{section}.\arabic{equation}}

%%%%%%%%%%%%%%%%%%%%%%%%%%%%%%%%%%%%%%%%%%%%%%%%%%%%%%%%%%%%%%%%%%%%%%%%%%%%%%%%%%%%%%%%%%
\section{QCD sum rules for the quark-gluon wave functions at the origin}\label{App:A}
%%%%%%%%%%%%%%%%%%%%%%%%%%%%%%%%%%%%%%%%%%%%%%%%%%%%%%%%%%%%%%%%%%%%%%%%%%%%%%%%%%%%%%%%%%

The definitions of quark-gluon twist-4 nucleon DAs (\ref{BMRa}) \cite{Braun:2008ia}
can be rewritten in conventional Dirac bispinor notation as follows:
\begin{widetext}
\begin{eqnarray}
  \label{eq:BMRb}
  ig\epsilon^{ijk}\langle 0 |\Big[u^i_\downarrow(z_1)C\slashed{n} u^j_\uparrow(z_2)\Big]
   \gamma^\nu\slashed{n} d^l_\downarrow(z_3) F_{n\nu}^{kl}(z_4)| p \rangle
&=&
   \frac{1}{4}m_N p_+^2\slashed{n} N^\uparrow(p) \int [dx]_3 \, e^{-ip_+\sum x_iz_i} \Phi_4^g(x),
\nonumber\\
  ig\epsilon^{ijk}\langle 0 |\Big[u^i_\uparrow(z_1)C\slashed{n} u^l_\downarrow(z_2)\Big]
   \gamma^\nu\slashed{n} d^k_\downarrow(z_3) F_{n\nu}^{jl}(z_4)| p \rangle
&=&
   \frac{1}{4}m_N p_+^2\slashed{n} N^\uparrow(p) \int [dx]_3 \, e^{-ip_+\sum x_iz_i} \Psi_4^g(x),
\nonumber\\
  ig\epsilon^{ijk}\langle 0 |\Big[u^i_\downarrow(z_1)C\gamma_\nu\slashed{n} d^j_\downarrow(z_2)\Big]
   \slashed{n} u^l_\downarrow(z_3) F_{n\nu}^{kl}(z_4)| p \rangle
&=&
   \frac{1}{4}m_N p_+^2\slashed{n} N^\uparrow(p) \int [dx]_3 \, e^{-ip_+\sum x_iz_i} \Xi_4^g(x),
\end{eqnarray}
\end{widetext}
where $F_{n\nu}^{kl} \equiv n^\mu F^a_{\mu\nu}(t^a)_{kl}$.
The normalization of the DAs is determined by the matrix elements of the corresponding
local operators. In what follows we estimate these matrix elements using the classical
SVZ QCD sum rule approach \cite{Shifman:1978bx}.

To this end we define isospin-1/2 twist-4 quark-gluon operators:
\begin{widetext}
\begin{eqnarray}
  \label{eq:GluonCurrents}
  \eta_1^{g}(x) &=&\frac{2}{3}ig\epsilon^{ijk}\Big[
       \Big(u^i(x)C\slashed{n} u^j(x)\Big)\gamma^\nu\slashed{n} d^l(x)
      -\Big(u^i(x)C\slashed{n} d^j(x)\Big)\gamma^\nu\slashed{n} u^l(x)\Big]F_{n\nu}^{kl}(x)\,,
\nonumber\\
  \eta_2^{g}(x) &=&\frac{2}{3}ig\epsilon^{ijk}\Big[
       \Big(u^i(x)C\gamma_5\slashed{n} u^l(x)\Big)\gamma^\nu\slashed{n} d^k(x)
      -\Big(u^i(x)C\gamma_5\slashed{n} d^l(x)\Big)\gamma^\nu\slashed{n} u^k(x)\Big]F_{n\nu}^{jl}(x)\,,
\nonumber\\
  \eta_3^{g}(x) &=&\frac{2}{3}ig\epsilon^{ijk}\Big[
       \Big(u^i(x)C\gamma^\nu\slashed{n} u^j(x)\Big)\slashed{n} d^l(x)
      -\Big(u^i(x)C\gamma^\nu\slashed{n} d^j(x)\Big)\slashed{n} u^l(x)\Big] F_{n\nu}^{kl}(x).
\end{eqnarray}
\end{widetext}
Matrix elements of these operators sandwiched between vacuum and the proton state
are related to the couplings introduced in Eq.~(\ref{eq:lambdag}):
\begin{eqnarray}
  \label{eq:LambdaGluon}
  \langle 0 |\eta_1^g(0)|p \rangle
%       &=& \frac{1}{2}\lambda_\Phi^g\,m_N p_+^2\slashed{n}\gamma_5N(p)\,,
       &=& -\frac{1}{4}\Big(\lambda_2^g-\frac13\lambda_3^g\Big)\,m_N p_+^2\slashed{n}\gamma_5N(p)\,,
\nonumber\\
  \langle 0 |\eta_2^g(0)|p \rangle
        &=& \phantom{-}\frac{1}{6}\Big(\lambda_2^{g}+\lambda_3^{g}\Big) m_N p_+^2\slashed{n} N(p)\,,
%       &=& \frac{1}{3}\Big(\lambda_\Phi^\text{g}+2\lambda_\Psi^\text{g}\Big) m_N p_+^2\slashed{n} N(p)\,,
\nonumber\\
  \langle 0 |\eta_3^g(0)|p \rangle
%       &=& \frac{1}{2}\Big(\lambda_\Phi^g\!+\!\lambda_\Psi^g\!+\!\lambda_\Xi^g\Big)m_N p_+^2\slashed{n}\gamma_5 N(p)\,.
       &=& \phantom{-}\frac{1}{6}\Big(\lambda_1^g + \lambda_3^g\Big)m_N p_+^2\slashed{n}\gamma_5 N(p)\,.
\end{eqnarray}
The sum rules are derived for the correlation functions of $\eta_k^{g}(x)$ with the three-quark
operators \cite{Ioffe:1981kw,Chung:1981cc}
\begin{eqnarray}
\label{eq:interpolator}
  \eta_1(x) &=& \epsilon^{ijk}\left[u^i(x)C\gamma_\mu u^j(x)\right]\gamma_5\gamma^\mu d^k(x)\,,
\nonumber\\
  \eta_2(x) &=& \epsilon^{ijk}\left[u^i(x)C\sigma_{\mu\nu} u^j(x)\right]\gamma_5\sigma^{\mu\nu} d^k(x)\,.
\end{eqnarray}
The corresponding couplings are well known from numerous QCD sum rule calculations
\begin{eqnarray}
 \langle 0 |\eta_1(0)|p\rangle &=&\lambda_1 m_NN(p)\,,
\qquad \lambda_1 \simeq - 2.7\cdot 10^{-2}~\text{GeV}^2\,,
\nonumber\\
\langle 0 |\eta_2(0)|p\rangle &=&\lambda_2 m_NN(p)\,,
\qquad \lambda_2 \simeq \phantom{-} 5.4\cdot 10^{-2}~\text{GeV}^2\,
\nonumber\\
\end{eqnarray}
where the numbers correspond to leading-order QCD sum rule results
at the scale 1 GeV, see e.g.~\cite{Braun:2006hz}.

In particular, we consider the following correlation functions:
\begin{eqnarray}
\frac{i}{4}\Tr
\left[\gamma_5\!\!\int\! d^4x\, e^{ipx}\langle 0|\mathrm{T}\{\eta_1^g(x)\bar \eta_1(0)\} |0\rangle\right]
&=& p^3_+\Pi^g_1(p^2)\,,
\nonumber\\
\frac{i}{4}\Tr
\left[\int\! d^4x\, e^{ipx}\langle 0|\mathrm{T}\{\eta_2^g(x)\bar \eta_1(0)\} |0\rangle\right]
&=& p^3_+\Pi^g_2(p^2)\,,
\nonumber\\
\frac{i}{4}\Tr
\left[\gamma_5 \!\!\int\! d^4x\, e^{ipx}\langle 0|\mathrm{T}\{\eta_3^g(x)\bar \eta_2(0)\} |0\rangle\right]
&=& p^3_+\Pi^g_3(p^2)\,.
\nonumber\\
\label{A:CF}
\end{eqnarray}
\begin{figure}[t]
\begin{center}
\includegraphics[width=0.35\textwidth,clip=true]{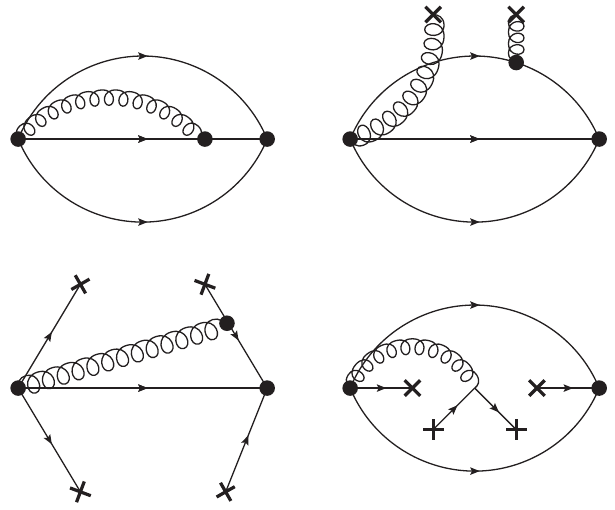}
\end{center}
\caption{
  Leading-order contributions to the OPE of the correlation functions in Eq.~(\ref{A:CF}).
        }
\label{fig:SR1}
\end{figure}
The sum rules are derived from the matching of the QCD calculation of
the invariant functions $\Pi^g_k(p^2)$ at Euclidean $p^2\sim -1$~GeV$^2$ with the dispersion
integral representation where the nucleon contribution is written explicitly:
\begin{eqnarray}
\Pi^g_1(p^2) &=&\phantom{-}\frac14 m_N^2 \frac{(\lambda_2^g-\lambda_3^g/3)\lambda_1}{m_N^2-p^2}+\ldots\,
\nonumber\\
\Pi^g_2(p^2) &=&\phantom{-}\frac16 m_N^2 \frac{(\lambda_2^g+\lambda_3^g)\lambda_1}{m_N^2-p^2}+\ldots\,
\nonumber\\
\Pi^g_2(p^2) &=&-\frac16 m_N^2 \frac{(\lambda_1^g+\lambda_3^g)\lambda_2}{m_N^2-p^2}+\ldots
\end{eqnarray}
and the
contributions of higher states and the continuum are modelled in the usual way as the
QCD spectral density above a certain threshold, $\sqrt{s_0}\sim 1.5$~GeV, dubbed the
interval of duality. On the QCD side, we take into account contributions of perturbation
theory and vacuum condensates of dimension 4 and 6 shown in Fig.~\ref{fig:SR1}.
The leading-order contributions of dimension 8 vanish for all cases.

\begin{figure}[t]
\begin{center}
\includegraphics[width=0.45\textwidth,clip=true]{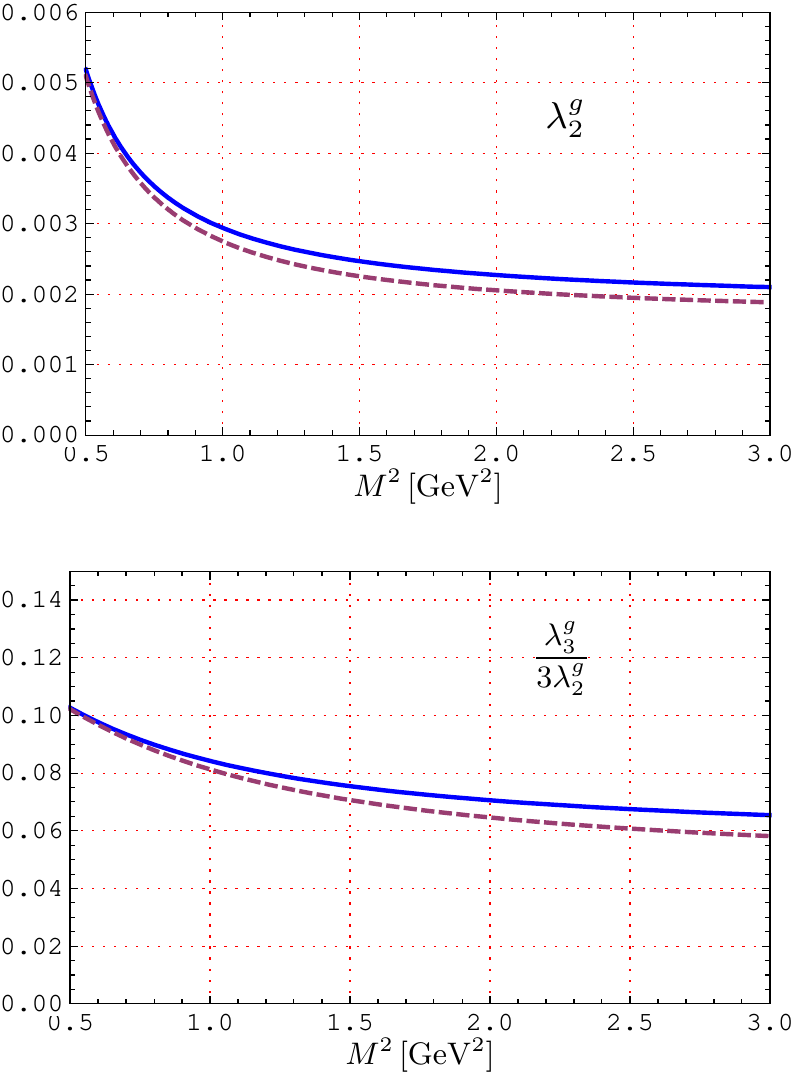}
\end{center}
\caption{ The coupling $\lambda_2^g$ (in units of $\text{GeV}^2$) (upper panel)
          and the ratio $\lambda_3^g/(3 \lambda_2^g)$ (lower panel)
          as a function of the
          Borel parameter $M^2$ for the central values of the condensates (\ref{eq:condensates}),
          $\alpha_s(1$~GeV)=0.5. The solid line corresponds to $\sqrt{s_0}=1.4$~GeV and
          the dashed line to $\sqrt{s_0}=1.6$~GeV. }
\label{fig:lambdag}
\end{figure}

Proceeding with the standard technique we derive the following set of sum rules:
\begin{eqnarray}
\label{eq:SumRule}
\lefteqn{
2(2\pi)^4(\lambda_2^g-\lambda_3^g/3)\lambda_1 m_N^2 e^{-\frac{m_{N}^2}{M^2}}=}
\nonumber\\&&{}\hspace*{1cm}=
  - \frac{\alpha_{s}}{45\pi}M^6 E_3
  - \frac{b}{12}M^2E_1 - \frac{8\alpha_{s}}{9\pi}a^2,
\nonumber\\
\lefteqn{
2(2\pi)^4(\lambda_2^{g}+\lambda_3^{g})\lambda_1 m_N^2 \, e^{-\frac{m_N^2}{M^2}} =}
\nonumber\\&&{}\hspace*{1cm}=
-\frac{\alpha_{s}}{45\pi} M^6 E_3 - \frac{b}{12}M^2E_1
 - \frac{40\alpha_{s}}{27\pi}a^2,
\nonumber\\
\lefteqn{
 2(2\pi)^4(\lambda_1^{g}+\lambda_3^{g})\lambda_2 m_N^2\, e^{-\frac{m_N^2}{M^2}}=}
\nonumber\\ &&{}\hspace*{1cm}=
 \frac{\alpha_{s}}{15\pi}M^6E_3 + \frac{b}{4}M^2E_1 + \frac{8\alpha_{s}}{3\pi}a^2,
\end{eqnarray}
where $M^2$ is the Borel parameter,
\begin{equation}
  \label{eq: En}
  E_n = 1-e^{-\frac{s_0}{M^2}}\sum\limits_{k=0}^{n-1}\frac{1}{k!}\left(\frac{s_0}{M^2}\right)^k
\end{equation}
and
\begin{align}
  \label{eq:condensates}
  a &=-(2\pi)^2\langle \bar qq\rangle =\left(0.55\pm 0.06\right)~\text{GeV}^3,
\nonumber\\
  b &= %(2\pi)^2\left\langle\frac{\alpha_{s}}{\pi}F^2\right\rangle = \left(0.47\pm 0.14\right)~\text{GeV}^4
 4\pi\langle \alpha_s\,F^2\rangle = \left(0.47\pm 0.14\right)~\text{GeV}^4.
\end{align}
are the quark and gluon condensates, respectively, at the scale 1 GeV.

For the numerical analysis we substitute the coupling $\lambda_1$ in the first two
equations
in~(\ref{eq:SumRule} by the square root of the ``Ioffe sum rule''~\cite{Ioffe:1981kw}
\begin{equation}
  \label{eq:IoffeSR}
  2(2\pi)^4 |\lambda_1|^2 m_N^2 e^{-\frac{m_N^2}{M^2}} = M^6E_3 + \frac{b}{4}M^2E_1 +
 \frac{a^2}{3}\Big(4-\frac{4}{3}\frac{m_0^2}{M^2}\Big),
\end{equation}
where $m_0^2 = \langle \bar q g\sigma F q\rangle/\langle \bar q q \rangle \simeq 0.65$~GeV$^2$,
and taking into account that $\lambda_1$ is negative (which is a convention).
Assuming the ``working window'' in the Borel parameter $M^2\sim 1-2$~GeV$^2$ and taking into account
uncertainties in the vacuum condensates and the continuum threshold $\sqrt{s_0}=1.4-1.6$~GeV, we
obtain the numbers given in Eq.~(\ref{eq:lambdag}) in the text.
Taking into account that to a good accuracy
$\lambda_2  = - 2\lambda_1$, we get $\frac23\lambda_1^g = \lambda_2^g-\lambda_3^g$.
This relation holds to the approximation considered here (leading-order QCD sum rules)
independent on the values of vacuum condensates and other parameters.
It can, however,  only be valid on a certain (low) normalization
scale as the anomalous dimensions of the couplings are different, cf.~(\ref{eq:gluon_adim}).

%%%%%%%%%%%%%%%%%%%%%%%%%%%%%%%%%%%%%%%%%%%%%%%%%%%%%%%%%%%%%%%%%%%%%%%%%%%%%%%%%%%%%%%%%%%%%%%
%%%%%%%%%%%%%%%%%%%%%%%%%%%%%%%%%%%%%%%%%%%%%%%%%%%%%%%%%%%%%%%%%%%%%%%%%%%%%%%%%%%%%%%%%%%%%%%

\end{document}